\documentclass[prd,aps,eqsecnum,floatfix,nofootinbib,preprint,tightenlines]{revtex4-2}

\usepackage{latexsym}
\usepackage{graphicx}
\usepackage{psfrag}
\usepackage[dvipsnames]{xcolor}
\usepackage{float}

\def\floatcaption#1#2{ \caption{#2 \label{#1}} }

\def\bibi{\bibitem}

\let\inodot=\i

\def\a{\alpha}
\def\b{\beta}
\def\c{\chi}
\def\d{\delta}
\def\g{\gamma}

\def\i{\iota}

\def\m{\mu}
\def\n{\nu}
\def\o{\omega}
\def\p{\pi}                     
\def\r{\rho}                    
\def\s{\sigma}                  
\def\t{\tau}

\def\D{\Delta}

\def\P{\Pi}

\def\cc{{\cal C}}

\def\cf{{\cal F}}

\def\co{{\cal O}}

\def\cbo{{\,\raise-.15ex\Sc [\,}}                       

\def\ddt#1{{\buildrel {\hbox{\LARGE .\kern-2pt.}} \over {#1}}}

\def\ie{\mbox{\it i.e.}}

\def\etc{\mbox{\it etc.}}

\def\half{{1\over 2}}

\def\ttl#1{{\it #1}}

\begin{document}

\vspace{-2cm}
\begin{center}
\begin{boldmath}
{\Large \bf The strong coupling from  hadronic $\t$-decay data\\
including $\t\to\p^-\p^0\n_\t$ from Belle}\\[1cm]
\end{boldmath}

Diogo Boito,$^a$ Aaron Eiben,$^b$ Maarten Golterman,$^b$
Kim Maltman,$^{c,d}$\\
Lucas M. Mansur,$^a$ and Santiago Peris$^e$\\[1cm]

$^a$Instituto de F{\'\inodot}sica de S\~ao Carlos, Universidade de S\~ao Paulo,
CP 369, 13560-970, S\~ao Carlos, SP, Brazil\\
[5mm]
$^b$Department of Physics and Astronomy, San Francisco State University,\\
San Francisco, CA 94132, USA\\
[5mm]
$^c$Department of Mathematics and Statistics,
York University\\  Toronto, ON Canada M3J~1P3
\\[5mm]
$^d$CSSM, University of Adelaide, Adelaide, SA~5005 Australia
\\[5mm]
$^e$Department of Physics and IFAE-BIST, Universitat Aut\`onoma de Barcelona\\
E-08193 Bellaterra, Barcelona, Spain
\\[10mm]

\end{center}

\begin{quotation}
In previous work we have combined the $\p^-\p^0$, $2\p^-\p^+\p^0$ and
$\p^-3\p^0$ spectral data obtained from hadronic $\t$ decays measured by
the ALEPH and OPAL experiments, together with electroproduction data
for several of the subleading hadronic modes and BaBar data for the
$K\bar{K}$ mode to construct an inclusive non-strange vector spectral
function entirely based on experimental data, with no Monte-Carlo
generated input. In this paper, we include, for the first time, the
Belle $\t\to\p^-\p^0\n_\t$ high-statistics decay data to 
construct a new inclusive non-strange vector spectral function that 
combines more of the world's available data. As no Belle data 
are at present available for the two $4\p$ modes, this requires a 
revised data analysis in comparison with our previous work.
From the resulting new spectral function, we obtain a new determination of
the strong coupling, $\alpha_s$, using our previously developed strategy
based on finite-energy sum rules. We find, at the $Z$ mass scale,
$\a_s(m_Z^2)=0.1159(14)$.  We discuss the smaller central value and larger 
error of our new result compared to our previous result,
showing the shifts to be due mainly to significant changes in updated HFLAV results for 
the $\p^-3\p^0$ decay mode.
\end{quotation}

\newpage
\section{\label{intro} Introduction}
There has been a longstanding interest in the determination of the strong
coupling, $\a_s$, from hadronic $\t$ decays. There are two main reasons
for this: first, the value obtained from $\t$ decays is among the more
precise ones \cite{PDG}, and second, it provides a valuable test of our
understanding of its running as predicted by QCD because all other
precise determinations have been obtained at considerably larger 
scales.

In this paper, we will be concerned with the determination of $\a_s$
from $\t$ decays in the non-strange vector ($V$) channel. There are
several experiments that have obtained data for hadronic $\t$ decays. The
LEP experiments ALEPH \cite{ALEPH,ALEPH2,ALEPH13} and OPAL \cite{OPAL}
produced results for the fully inclusive non-strange (isovector) $V$
spectral function obtained by supplementing measured contributions from the
dominant decay modes, $\p^-\p^0$, $2\p^-\p^+\p^0$ and $\p^-3\p^0$, with
those from a number of additional exclusive modes with small branching
fractions (BFs) estimated using Monte-Carlo simulations. More recently,
the KEK-based Belle experiment \cite{Belle} has produced results for
the contribution to the spectral function of the $\p^-\p^0$ exclusive
mode, which has the largest of the $V$ channel hadronic BFs. Information 
on this mode is also available from CLEO \cite{CLEO,CLEO2pi}.

Clearly, a robust determination of $\a_s$ should make use of the world's
data on hadronic $\t$ decays, after investigating whether the available
data sets are statistically compatible. In previous work \cite{alphas20}
we combined the ALEPH and OPAL data for the $\p^-\p^0$, $2\p^-\p^+\p^0$
and $\p^-3\p^0$ modes (which we will refer to as the $2\p+4\p$ data)
into a three-mode-summed $2\pi +4\pi$ spectral function contribution.
By BF, these three modes constitute about 98\% of the inclusive
non-strange $V$ channel contribution to hadronic $\t$ decays. We then
used for the $K\bar{K}$ spectral contributions recent BaBar differential 
$\t\to K\bar{K}\n_\t$ \cite{babarkkbartau18} distribution results,
and for the contributions from the remaining modes,
$\p^- \o (\rightarrow \mbox{non-}3\p)$ \cite{sndomegapi16,babaromegapi17},
$\eta \p^-\p^0$ \cite{Belleetapipi09,sndetapipi1518,sndetapipi15182,babaretapipi18,cmd3etapipi19,cmd3etapipicovs},
$K\bar{K}\p$ \cite{babarepemkkbarpi},
$6\p$ \cite{babarsigma6pi06,thankssolodov,cmd3sigma3pim3pip13,cmd3sigmaeta3pi17,sndsigmapimpip4pi019,
sndsigmaeta3pi19},
$K\bar{K}\p\p$ \cite{babarepemkkbarpi},
$(3\p)^-\o (\rightarrow {\rm \mbox{non-}3}\p)$ \cite{babarsigma6pi06},
$\p^-\eta\o(\rightarrow {\rm \mbox{non-}3}\p)$ \cite{babarepemetapimpip2pi018} and
$\eta 4\p$ \cite{babarepemetapimpip2pi018,babarepemeta2pim2pip07},
results obtained using electroproduction cross section data in
combination with the CVC (Conserved Vector Current)
relation.\footnote{We follow Ref.~\cite{alphas20} in
referring to exclusive modes other than $2\pi$ and $4\pi$ as ``residual
modes.'' See Ref.~\cite{alphas20} for the complete analysis of these
residual-mode contributions.}  
The contributions from these modes had previously been
estimated using input from Monte-Carlo simulations.
Exclusive-mode $\tau$-decay BFs were obtained from Ref.~\cite{hflav2019}
where needed. These additions produced an inclusive non-strange $V$
channel spectral function with experimental data used for exclusive-mode
contributions representing at least 99.95\% by BF of the inclusive non-strange
$V$ total, removing the need for Monte-Carlo input.

Because of significantly increased statistics relative to the LEP-based
experiments, the errors on the unit-normalized Belle
$\p^-\p^0$ data are much smaller than those of ALEPH and OPAL, and it
thus makes sense to combine the $\p^-\p^0$ data of all three experiments
with the aim of increasing the overall precision of the non-strange $V$
inclusive spectral function. Including the Belle data is thus 
one of the main goals of this paper. The possibility of including 
the older CLEO $\p^-\p^0$ data in this combination is also considered,
though this is hampered by the lack of complete information on the CLEO
systematic errors.

The inclusion of the Belle $\p^-\p^0$ data requires a complete 
redesign of the algorithm employed in our previous $2\p+4\p$ data combination
based on the ALEPH and OPAL data. The reason is that in Ref.~\cite{alphas20} we
first added the $\p^-\p^0$ and two $4\p$ contributions for each of
ALEPH and OPAL separately (using BFs from Ref.~\cite{hflav2019}) before
forming the two-experiment, three-mode sum, instead of first combining
ALEPH and OPAL contributions for each of the three modes separately and
then combining those results into the two-experiment, three-mode sum. The
reason for this choice was that the covariance matrices of the separate
ALEPH and OPAL $2\p^-\p^+\p^0$ and $\p^-3\p^0$ contributions to the
unit-normalized exclusive-mode spectral function are poorly determined,
but adding the three contributions for each experiment first made the
combined $2\p+4\p$ covariance matrices well-behaved for both ALEPH
and OPAL, thus allowing us to combine the two
$2\pi +4\pi$ data sets.\footnote{The three-mode sum
is dominated by $\p^-\p^0$ contributions, for which the
ALEPH and OPAL covariance matrices are well known and well behaved.}
Clearly, this procedure needs to be revised when the Belle 
$\p^-\p^0$ data are included in the combination as well.

The modification of our strategy to allow for the inclusion of the 
Belle  $\p^-\p^0$ data set, also, in principle, allows us to include 
the older CLEO $\p^-\p^0$ data set \cite{CLEO,CLEO2pi}. Although CLEO 
provided a careful estimate of systematic errors for integrated quantities such as
$a_\m^{\rm HVP}$, the hadronic vacuum polarization contribution to the 
muon anomalous magnetic moment, a covariance matrix for the systematic 
errors bin-by-bin was not provided.\footnote{We thank Jon Urheim for an 
email exchange on this topic.} We have, nonetheless, also investigated the 
effect of including the CLEO $\pi^- \pi^0$ data, verifying its consistency 
with the results from the other experiments. Since, however, the information 
on the CLEO systematic uncertainties is less complete than that for the 
other three experiments, the quantitative results of this exploration are 
not taken into account in obtaining the main results of this paper.

The data combination we performed in our previous work \cite{alphas20}
employed the algorithm of Ref.~\cite{KNT18}, in which the experimental data
is divided among a number of ``clusters.'' The combined spectral function
is then defined by interpolation between the cluster values, which,
in turn, are fitted to the experimental data. A detailed
description of this procedure is provided in Sec.~\ref{algorithm}.

The strategy we adopt here can be summarized as follows. We first carry out
the $\p^-\p^0$ combination using ALEPH, OPAL and Belle
unit-normalized data,
following the algorithm of Ref.~\cite{KNT18}, and then multiply the result
with the BF from the updated HFLAV analysis \cite{hflav22}. We then
consider the ALEPH and OPAL $4\p$ data in more detail. The covariance
matrix for the OPAL $\p^-3\p^0$ data is particularly badly behaved,
because of large backgrounds (in particular from other modes with one
$\p^-$, such as $\p^-\p^0$ and $\p^-2\p^0$) making it difficult to reliably
determine the $\p^-3\p^0$ covariances \cite{OPAL}.\footnote{We thank
Sven Menke for an email exchange on this topic.} To deal with this
problem, we include only the diagonal entries of the OPAL $\pi^- 3\pi^0$
covariance matrix in our fit combining the
ALEPH and OPAL $4\p$ data, while using the full covariance matrix for error
propagation. Doing so, we found that, if we combine the two $4\p$
modes for each of ALEPH and OPAL first, we can carry out a pure-$4\p$
fit to these data, thus obtaining a combined ALEPH plus OPAL two-mode,
$4\p$ spectral function contribution. Finally, we add the combined
$2\p$ and combined $4\p$ spectral function contributions ({\it c.f.} 
Sec.~\ref{full}). While correlations between different modes have not 
been provided by the ALEPH and OPAL experiments, we take into account 
the small correlations induced by the correlations between the BFs of 
the three modes as provided by HFLAV \cite{hflav22}.

The contribution from the residual modes will be taken in unmodified form
from Ref.~\cite{alphas20} because any more recent updates of $\tau$ BFs
and/or the relevant exclusive-mode electroproduction cross sections do
not affect our results within errors. We remind the reader that isospin
breaking associated with the use of CVC for contributions obtained
using electroproduction data and CVC are too small to have any
effect, since contributions from these modes lie at squared invariant
masses well above the $\r-\o$ resonance region. Isospin-breaking
corrections to these contributions will thus receive no narrow, nearby
$I=0/1$ resonance interference enhancements, and hence are expected
to be at the $O(1\% )$ (or less) level, much smaller than the errors on
the CVC-converted contributions. Adding the residual-mode and $2\pi +4\pi$
contributions we obtain a new inclusive non-strange, vector-isovector
spectral function which now includes the Belle $\pi^-\pi^0$ data.

In the second part of this paper, we use finite-energy sum rules
(FESRs) to fit $\a_s$ to moments of the new non-strange $V$ spectral
function. To do this, we follow the strategy of Ref.~\cite{alphas20},
comparing our results with those obtained in that earlier work. In
this strategy, non-perturbative effects are taken into account through the
operator product expansion (OPE) and a model for quark-hadron duality
violations (DVs). This strategy is fundamentally different from the alternate
``truncated OPE'' (tOPE) approach used in a number of earlier analyses
in the literature Refs.~\cite{ALEPH,ALEPH2,ALEPH13,OPAL,DDHMZ08,PT,Pich,PRS}.
Within the tOPE, $\a_s$ turns out to be
determined from FESRs involving high-degree polynomial weights in which only
perturbative contributions are retained in the corresponding theoretical
representations \cite{DVvstOPE}. The neglect of in-principle-present
non-perturbative (NP) contributions (in particular those proportional to
higher-dimension OPE condensates) is potentially dangerous at scales as low
as the $\t$ mass. In Ref.~\cite{DVvstOPE}, for example, it was shown that a $V$-channel 
optimal-weight tOPE analysis at a scale $s_0=2.88$ GeV$^2$ (which is 
slightly higher than that employed in the ALEPH-based $V$-channel tOPE 
analyses of Refs.~\cite{Pich,PRS}), based on the improved $V$ spectral function 
of Ref.~\cite{alphas20}, produced a discrepancy larger than $6\s$ between the two
single-weight $\a_s$ determinations obtained from the underlying NP-free 
FESRs, leading to a very large $\c^2/{\rm dof} \simeq 43/1$ for the fit
combining these two NP-free FESRs.  Since similarly large $\c^2$ are found 
for $s_0\simeq 2.88$~GeV$^2$ tOPE analyses of the types employed in 
Refs.~\cite{Pich,PRS} employing the improved $V$-channel spectral function
constructed in the present paper, we do not consider further the tOPE 
approach in what follows.\footnote{We refer to 
Refs.~\cite{critical,EManalysis,DVvstOPE} for a more detailed 
assessment of the tOPE approach.}

In the (dominant) perturbative part of our FESR-based analysis,
we use the results of Ref.~\cite{PT} in the standard $\overline{\rm MS}$
scheme, which, in the context of FESRs, is usually referred to as
fixed-order perturbation theory (FOPT). The alternative approach
based on a resummation known as contour-improved perturbation theory (CIPT)
\cite{CIPT,CIPT2} was recently shown to be inconsistent with the standard OPE
\cite{HR,GMP,GHM,BH1,BH2} and will thus not be employed.

This paper is organized as follows. In Sec.~\ref{theory} we briefly summarize
the FESRs that allow us to extract $\a_s$ from the inclusive non-strange
$V$ spectral function. Section~\ref{datasec} is devoted to the construction
of the new spectral function. In particular, in this section we explain our
strategy for combining the ALEPH and OPAL $4\p$ data, and adding this
to the combined $2\p$ data and the residual-mode data from Ref.~\cite{alphas20} 
to arrive at a new inclusive result that now also includes the $\p^-\p^0$ 
data from Belle. In Sec.~\ref{strong
coupling}, we use the framework of Sec.~\ref{theory} to produce our new result
for $\a_s$ at the $\t$ mass. Section~\ref{conclusion} contains our 
conclusions. There are three appendices.   The first details our handling of the d'Agostini 
bias \cite{Abias} in the $4\p$ data combination step, and the second one summarizes a 
study of the effects of including the CLEO $\p^-\p^0$ data set in 
addition to those of ALEPH, OPAL and Belle. The third one contains details
on combining the $2\p$ and $4\p$ spectral functions.

\section{\label{theory} Theory review}
Our theoretical framework for the determination of $\a_s(m_\t^2)$ from the
inclusive non-strange $V$ spectral function is the same as that of
Ref.~\cite{alphas20}, and we will therefore be very brief. More details can also
be found in earlier work, notably in the first paper in which we employed
this strategy, Ref.~\cite{alphas1}, but see also Refs.~\cite{alphas2,alphas14}.

\subsection{\label{FESR} Finite energy sum rules}
As we will only consider the non-strange $V$ channel in this paper,
we will limit our review to that case. The key object entering the
analysis of this channel is the spin-$1$ scalar vacuum polarization,
$\Pi^{(1)}$, of the isovector, $V$ current-current two-point function.
Neglecting the very small up-down quark mass difference, the two-point
function is purely transverse and related to $\Pi^{(1)}$ by
\begin{equation}
\label{correl}
\P_{\m\n}(q)=i\int d^4x\,e^{iqx}\langle 0|T\left\{J_\m(x)
J^\dagger_\n(0)\right\}|0\rangle=\left(q_\m q_\n-q^2 g_{\m\n}\right)
\P^{(1)}(q^2)\ ,\nonumber
\end{equation}
where $J_\m$ is the isovector $V$ current $\overline{u}\g_\m d$.
$\Pi^{(1)}$ is also dispersively related to the corresponding
spin-$1$ spectral function,
\begin{equation}
\label{spectral}
\r^{(1)}(s)=\frac{1}{\p}\;\mbox{Im}\,\P^{(1)}(s)\ ,
\end{equation}
with $s=q^2$. $\P^{(1)}$ and $\rho^{(1)}$ satisfy the Cauchy
Theorem (FESR) relation,
\begin{equation}
\label{cauchy}
I^{(w)}(s_0)\equiv\frac{1}{s_0}\int_0^{s_0}ds\,w(s)\,\r^{(1)}(s)
=-\frac{1}{2\p i\, s_0}\oint_{|z|=s_0}
dz\,w(z)\,\P^{(1)}(z)\ ,
\end{equation}
which is valid for any $s_0>0$ and any weight $w(s)$
analytic inside and on the contour $|z|=s_0$ in the complex plane
\cite{shankar,MPR,CK,CKT,KPT,FNR,BLR,Braaten88,BNP}.\footnote{If
one retains the up-down quark mass difference, the two-point function also
depends on a scalar, spin-$0$ polarization, whose spectral function is
$O((m_d-m_u)^2)$, and which is safely negligible given the precision of current
isovector $V$ spectral data. With spin $0$ contributions negligible, we
will drop the superscripts $(1)$ in what follows and denote $\Pi^{(1)}$
by $\Pi$ and $\rho^{(1)}$ by either $\rho$ or $\rho_{ud;V}$.}
We will choose $w(z)$ to be polynomial in $z$.

The basic idea for extracting $\alpha_s$ from hadronic $\tau$ decay
data is then to evaluate the left-hand side of Eq.~(\ref{cauchy}) using
experimental data for the spectral function, while taking $s_0$ large
enough that QCD perturbation theory provides a good representation of the
right-hand side. As the value of $s_0$ for which spectral data is
available is kinematically limited to $s_0\leq m_\t^2$, non-perturbative
corrections to the right-hand side are potentially non-negligible,
and we discuss our framework for including them below. When the
need arises, we will refer to the $w(s)$-weighted spectral integral
above as $I_{\rm ex}^{(w)}(s_0)$, and the corresponding weighted
$|z|=s_0$ contour integral as $I_{\rm th}^{(w)}(s_0)$.

For large enough $|s|=s_0$, and sufficiently far away from the Minkowski
axis $z=s>0$, $\P(s)$ can be approximated by the OPE
\begin{equation}
\label{OPE}
\P_{\rm OPE}(z)=\sum_{k=0}^\infty \frac{C_{2k}(z)}{(-z)^{k}}\ ,
\end{equation}
where the logarithmic $z$ dependence of the OPE coefficients $C_{2k}$ can
in principle be calculated in perturbation theory.

For the $k=0$ term, one usually considers, instead of $\P(z)$, the Adler
function $D(z) \equiv -z\,d\P(z)/dz$, which is finite and formally independent
of the renormalization scale $\m$. Accordingly, the $k=0$ contribution to
the right-hand side of Eq.~(\ref{cauchy}) can be expressed in terms of 
the corresponding contribution to the Adler function via partial integration.
The dimension $D=0$ contribution $D_0(z)$ to $D(z)$ takes the form
\begin{equation}
\label{pertth}
D_0(z)\equiv -z\,\frac{dC_0(z)}{dz}=\frac{1}{4\p^2}\sum_{n=0}^\infty
\left(\frac{\a_s(\mu^2)}{\p}\right)^n\sum_{m=1}^{n+1}
mc_{nm}\left(\log\frac{-z}{\m^2}\right)^{m-1}\ ,
\end{equation}
where the coefficients $c_{nm}$ are known to order $\a_s^4$~\cite{PT}.
In the $\overline{\rm MS}$ scheme, $c_{01}=c_{11}=1$, $c_{21}=1.63982$,
$c_{31}=6.37101$ and $c_{41}=49.07570$, for three flavors \cite{PT}; all
other coefficients to this order can be determined using the scale
independence of $D_0(z)$ and the renormalization group. While $c_{51}$ is
not known at present, we will use the estimate $c_{51}=283\pm 142$, which
generously covers all estimates for this coefficient that can be found in
the literature~\cite{PT,BJ,BMO,IC}. For the running of $\a_s$ we use the
four-loop $\overline{\rm MS}$ $\b$-function, but we have checked that
using five-loop running instead \cite{5loop,5loop2} leads to differences
of order $10^{-4}$ or less in our results for $\a_s(m_\t^2)$. As already
mentioned in the introduction, for evaluating the perturbative contribution
to the FESR~(\ref{cauchy}) we will choose $\m^2=s_0$ in Eq.~(\ref{pertth}), which
corresponds to using FOPT.

The $C_{2k}$ contain non-perturbative $D=2k$ condensate
contributions for $k>1$. As in Ref.~\cite{alphas20}, we will neglect purely
perturbative quark-mass contributions to $C_{2k}$, $k\ge 1$, as they
are numerically very small for the non-strange $V$ channel
FERSs considered in this paper. We will also neglect the $z$-dependence
of the coefficients $C_{2k}$ for $k>1$.\footnote{Such logarithmic
corrections, which are suppressed by additional powers of $\alpha_s$,
are known only for low values of $k$.}
It then follows that, for weights $w(y)$ polynomial in $y=s/s_0$,
$D=2k>2$ contributions to the contour integral $I_{\rm th}^{(w)}(s_0)$
are proportional to $C_{2k}/s_0^k$, and present only if $w(y)$ contains
a term proportional to $y^{k-1}$. A more detailed discussion of
our treatment of the $D>0$ OPE contributions, including a response to
Ref.~\cite{PRS}, may be found in Refs.~\cite{DVvstOPE,alphas1}.

So far, our strategy is based on standard OPE-based QCD tools. However, the OPE
breaks down near the Minkowski axis \cite{PQW}. If the OPE would
also hold for $z=s>0$, Eq.~(\ref{cauchy}) would establish a direct
correspondence between the OPE and the resonant behavior of the
experimental spectral function, generally referred to as quark-hadron
duality. However, it is clear that the OPE cannot account for the
oscillatory resonance behavior of the spectral function seen up to fairly
large values of $s$, and thus quark-hadron duality is violated. We briefly 
review our approach to this problem, as this is where
our strategy differs from others used in the literature.

We account for the breakdown of this duality by
replacing the right-hand side of Eq.~(\ref{cauchy}) by
\begin{equation}
\label{split}
-\frac{1}{2\p is_0}\oint_{|z|=s_0}dz\,w(z)\,
\left(\P_{\rm OPE}(z)+\D(z)\right)\ ,
\end{equation}
with
\begin{equation}
\label{DVdef}
\D(z)\equiv\P(z)-\P_{\rm OPE}(z)\ ,
\end{equation}
defining the quark-hadron duality violating
contribution $\D(z)$ to $\Pi(z)$. One expects $\D(z)$ to decay
exponentially for $|z|\to\infty$, and thus for polynomial weights
Eq.~(\ref{split}) can be rewritten as \cite{CGP}
\begin{equation}
\label{sumrule}
I_{\rm th}^{(w)}(s_0) = -\frac{1}{2\p is_0}\oint_{|s|=s_0}
dz\,w(z)\,\P_{\rm OPE}(z)-\frac{1}{s_0}\,
\int_{s_0}^\infty ds\,w(s)\,\frac{1}{\p}\,\mbox{Im}\,
\D(s)\ .
\end{equation}
The imaginary part $\frac{1}{\p}\,\mbox{Im}\,\D(s)$ can be interpreted as
the duality-violating part $\rho^{\rm DV}(s)$ of the non-strange $V$
spectral function, and represents the resonance-induced, oscillatory
parts of the spectral function not captured by the OPE.

In Ref.~\cite{BCGMP}, we developed a theoretical framework for quark-hadron
duality violations in terms of a generalized Borel--Laplace transform of
$\P(q^2)$ and hyperasymptotics, building on earlier
work \cite{russians,russians2,russians3,catalans}. In the chiral limit,
and assuming the spectrum becomes Regge-like asymptotically at large
$s$ in the $N_c\to\infty$ limit, we showed that the large-$s$ form of
$\r^{\rm DV}(s)$ can be parametrized as
\begin{equation}
\label{ansatz}
\r^{\rm DV}(s)=\frac{1}{\p}\,\mbox{Im}\,
\D(s)=\left(1+\frac{c}{s}\right)e^{-\d-\g s}\sin(\a+\b s)\ ,
\end{equation}
up to slowly varying logarithmic corrections and with $\g\sim 1/N_c$ small but non-zero.\footnote{This
form (without the $c/s$ correction) was first introduced in Ref.~\cite{CGP05}, and subsequently used in
Refs.~\cite{alphas1,alphas2,alphas14,CGP,CGPmodel}.} The parameter $\b$ is
directly related to the Regge slope, and the parameter $\g$ to the
(asymptotic) ratio of the width and mass of the resonances. 
The DV parameters are to be fitted along with $\a_s(m_\t^2)$ and the other
OPE parameters to the weighted spectral integrals of Eq.~(\ref{cauchy}).
In Eq.~(\ref{ansatz}), we have allowed for a multiplicative correction of 
order $1/s$ \cite{BCGMP}, the effect
of which will be quantitatively investigated in Sec.~\ref{analysis}.
While the framework of Ref.~\cite{BCGMP} is rather general and based on
generally accepted conjectures about QCD (primarily Regge-like
behavior), Ref.~\cite{BCGMP} does not provide a first-principle derivation
from QCD. This introduces model dependence into our analysis which,
however, can be tested by fits to the data. 
Such tests, in particular, will provide information about the values
of $s$ above which this asymptotic form is likely to be sufficiently
accurate. We emphasize that modifications to the parametrization of
Eq.~(\ref{ansatz}) are constrained by the general framework of Ref.~\cite{BCGMP}.
For a more recent discussion of our framework we refer to Ref.~\cite{DVvstOPE},
which also addresses recent criticism leveled against our
strategy in Ref.~\cite{PRS}.

In summary, as in Refs.~\cite{alphas1,alphas2,alphas14,alphas20}, we will assume
that Eq.~(\ref{ansatz}) holds for $s\ge s_{\rm min}$, with $s_{\rm min}$ to be
determined from fits to the data. This assumes of course that the
$s_{\rm min}$ for which this assumption is valid includes a region
below $m_\t^2$, \ie, that both the OPE~(\ref{OPE}) and the DV
parametrization~(\ref{ansatz}) can be used in some interval below $m_\t^2$.

\subsection{\label{weightOPE} Choice of weight functions and the OPE}
The logarithmic $s$ dependence of the OPE coefficients $C_D(s)$ is
calculable in perturbation theory; this $s$ dependence is an $\co(\a_s^2)$
effect in the chiral limit. Such effects were found to be safely negligible
for $D=4$ and $D=6$ in the sum-rule analysis of the OPAL data reported in
Ref.~\cite{alphas1}, and again in Ref.~\cite{DVvstOPE}, and we will thus ignore
them for $D>0$ in the present analysis as well.\footnote{Nothing is known
about logarithmic corrections beyond $D=6$.} This assumption is
common to all $\alpha_s$ analyses using $\tau$-decay data. As noted
above, assuming the coefficients $C_D(z)$ to be $z$ independent, a term
proportional to the monomial $z^n$ in the (polynomial) weight $w(z)$
projects onto the $D=2(n+1)$ OPE contribution in the sum
rule~(\ref{cauchy}).\footnote{The $D=0$ term, perturbation theory,
of course contributes for all $n$.} If $w(z)$ has degree $k$, the right-hand
side of ~(\ref{cauchy}) then contains only a finite number of $D>0$ OPE
contributions, having maximum dimension $D=2k+2$.

In this paper, we will consider the weights $w(z)=w_n(z/s_0)$ with
\begin{eqnarray}
\label{weights}
w_0(y)&=&1\ ,\\
w_2(y)&=&1-y^2\ ,\nonumber\\
w_3(y)&=&(1-y)^2(1+2y)\ ,\nonumber\\
w_4(y)&=&(1-y^2)^2\ ,\nonumber
\end{eqnarray}
where the subscript indicates the degree of the polynomial. These weights
explore OPE terms with $D\le 10$, and form a linearly independent basis
for polynomials up to degree four without a linear term. The weight
$w_0(y)$ projects only the $D=0$ term of the OPE (\ie, pure perturbation
theory), while the weight $w_2(y)$ projects onto $D=0$ and $D=6$,
the weight $w_3(y)$ onto $D=0$, $D=6$ and $D=8$, and the weight $w_4(y)$
onto $D=0$, $D=6$ and $D=10$. As the OPE is at best an asymptotic
expansion in $1/z$, it is safer to limit oneself to sum rules with
low-degree weights such as $w_0(y)$ and $w_2(y)$, and check for
consistency among sum rules with different, higher-degree weights.
As explained in Ref.~\cite{alphas1}, we choose to avoid weights with a
linear term in $y$, as renormalon-based studies suggest that such weights
are perturbatively unstable \cite{BJ,MJ,BBJ12,BO20}.

The different sensitivity of these weights to DVs also allows us to
have some control on the possible systematic error coming from the
approximate nature of the DV parametrization~(\ref{ansatz}).
The weights $w_{2,3,4}(y)$ are ``pinched,'' \ie, they have zeros at
$z=s_0$, and thus suppress contributions from the region near the
timelike point $z=s_0$ on the contour and hence also DV
contributions to the associated FESRs \cite{KM98,DS99}. The weight 
$w_2(y)$ has a single zero at $z=s_0$ (a single pinch), while the weights 
$w_3(y)$ and $w_4(y)$ are doubly pinched, \ie, have a double zero at 
$z=s_0$.\footnote{The oscillatory dependence of $\rho^{\rm DV}(s)$ on 
$s$, of course, produces an oscillatory dependence of weighted DV integrals 
on $s_0$. Employing FESRs with multiple $s_0$ thus provides implicit tests 
of the assumed DV form.} In order to maintain maximal sensitivity to 
the DV parameters $\d$, $\g$, $\a$ and $\b$, we will always include the 
$w_0(y)$ FESR in our fits. For a more detailed discussion of this
aspect of our strategy, we refer to Sec.~\ref{strategy}.

\section{\label{datasec} Data}

In this section, we will describe in detail our procedure for
combining the ALEPH, OPAL and Belle $\p^-\p^0$ data
and ALEPH and OPAL $2\p^-\p^+\p^0$ and $\p^-3\p^0$ data.
We begin by detailing the sources of all the data used in this paper
in Sec.~\ref{inputs}, after which, in Sec.~\ref{algorithm}, we describe
our data-combination algorithm. We then present the $\p^-\p^0$
combination in Sec.~\ref{2picomb} and the two-mode $4\p$ combination
of ALEPH and OPAL $2\p^-\p^+\p^0$ and $\p^-3\p^0$ data in
Sec.~\ref{4picomb}. In Sec.~\ref{full} the resulting $2\p$ and $4\p$
contributions are added together, using suitable interpolation and
taking into account correlations between the $2\p$ and $4\p$
contributions introduced by the BF correlations. Finally, in
Sec.~\ref{specfun}, we add the residual mode contributions
from Ref.~\cite{alphas20} to arrive at our new result for the inclusive
non-strange $V$ spectral function, $\rho (s)$. In what follows,
we will denote the contribution of a given exclusive mode, or exclusive-mode
combination, $X$, to $\rho (s)$ as $\rho_X(s)$, and refer to this quantity
as ``the mode-$X$ spectral function.''
As explained in Sec.~\ref{intro}, we do not include the CLEO data
for the $\p^-\p^0$ mode in the construction of the inclusive spectral 
function.   Instead, we discuss the impact of including the CLEO data in 
App.~\ref{CLEOsec}, where it is shown that the impact is small.

\subsection{\label{inputs} External input}

ALEPH exclusive-mode $X=\p^-\p^0,\ 2\p^-\p^+\p^0,\ \p^-3\p^0$ data
are provided in the form of the BF-normalized experimental number
distributions ${\frac{B_X}{N_X}}{\frac{dN_X(s)}{ds}}$. These are related
to the corresponding mode-$X$ spectral functions by \cite{ALEPH}
\begin{equation}
\rho_X(s) = {\frac{m_\tau^2}{12\pi^2 B_e S_{\rm EW}
\vert V_{ud}\vert^2 w^{\rm av}_T(s;m_\t^2)}} \,
{\frac{B_X}{N_X}}{\frac{dN_X(s)}{ds}}
\label{dBdstorho}
\end{equation}
with $B_X$ the $\tau\rightarrow X\nu_\tau$ BF for exclusive mode $X$,
$B_e$ the $\t \rightarrow e\n_\t \bar{\n}_e$ BF, and $w^{\rm av}_T(s;s_0)$
defined as \cite{alphas14}
\begin{equation}
\label{binav}
w^{\rm av}_T(s={\tt sbin(i)};m_\t^2)=\frac{1}{\tt dsbin(i)}
\int_{\tt sbin(i)-dsbin(i)/2}^{\tt sbin(i)+dsbin(i)/2}
w_T(s;m_\t^2)\,ds\ ,
\end{equation}
where {\tt sbin(i)} is the $i$-th bin center, {\tt dsbin(i)} is the $i$-th
bin width, and $w_T(s;s0)$ is the kinematic weight
\begin{equation}
\label{wT}
w_T(s;s_0)\equiv w_3(s/s_0)=\left(1-\frac{s}{s_0}\right)^2
\left(1+2\,\frac{s}{s_0}\right)\ .
\end{equation}
To perform the conversion~(\ref{dBdstorho}) we require the hadronic
BFs $B_X$, together with input for the external parameters, $B_e$,
the short-distance electroweak correction factor $S_{\rm EW}$, the $ud$
Cabibbo--Kobayashi--Maskawa matrix element $|V_{ud}|$ and the $\t$
mass $m_\t$. The external parameters appear in the fixed combination
\begin{equation}
\label{C}
\cf=\frac{12\pi^2 B_e S_{\rm EW}\vert V_{ud}\vert^2}{m_\t^2}\ .
\end{equation}
We employ for the BFs $B_X$ and correlations between them the HFLAV
2022 Ref.~\cite{hflav22} results\footnote{The BFs and errors in
Eqs.~(\ref{BX}) are unchanged in the even more recent HFLAV 2024 
update~\cite{hflav24}. HFLAV 2022, however, quotes the BFs and errors
to one more significant figure than does HFLAV 2024 and has the 
additional advantage of listing the basis-mode BF correlations, which 
are not tabulated in HFLAV 2024.}
\begin{eqnarray}
\label{BX}
B_{2\p}\equiv B_{\p^-\p^0}&=&0.25486(90) \ ,\\
B_{1\p^0}\equiv B_{2\p^-\p^+\p^0}&=&0.04480(55) \ ,\nonumber\\
B_{3\p^0}\equiv B_{\p^-3\p^0}&=&0.01040(71) \ ,\nonumber\\
r_{\p^-\p^0,2\p^-\p^+\p^0}&=&-0.13 \ ,\nonumber\\
r_{2\p^-\p^+\p^0,\p^-3\p^0}&=&-0.01 \ ,\nonumber\\
r_{\p^-\p^0,\p^-3\p^0}&=&+0.01\ ,\nonumber
\end{eqnarray}
where $r_{X,Y}$ is the correlation between $B_X$ and 
$B_Y$. Note that the 2022 HFLAV updates differ from the then-current,
HFLAV 2019 Ref.~\cite{hflav2019} inputs used in Ref.~\cite{alphas20}. Of
particular numerical significance are the 9\% downward shift in
central value and factor of $2.4$ increase in error relative to
HFLAV 2019 seen in the HFLAV 2022 $B_{\p^-3\p^0}$ result. The
values used for the external parameters (omitting the errors on $m_\t$,
$\vert V_{ud}\vert$, and $S_{\rm EW}$, for which the relative errors are much smaller than
that on $B_e$) are (from, repectively, Ref.~\cite{PDG} for $m_\t$ and $|V_{ud}|$, Ref.~\cite{hflav22}for $B_e$ and Ref.~\cite{erlersew} for $S_{\rm EW}$)
\begin{eqnarray}
\label{other}
m_\t&=&1.77693~\mbox{GeV}\ ,\\
|V_{ud}|&=&0.97367\ ,\nonumber\\
B_e&=&0.17812(22)\ ,\nonumber\\
S_{\rm EW}&=& 1.0201\ ,\nonumber
\end{eqnarray}
and lead to
\begin{equation}
\label{Cinv}
\cf=6.4613(80)~\mbox{GeV}^{-2}\ .
\end{equation}
The conversion relation Eq.~(\ref{dBdstorho}) and inputs above
are, of course, also used to convert the covariances of the ALEPH
BF-normalized number-distributions to those of $\rho_X(s)$.

OPAL quotes results in the form of the exclusive-mode contributions
$\rho_X(s)$. These were obtained from the experimentally measured
normalized number distributions via Eq.~(\ref{dBdstorho}), using
then-current values of the exclusive-mode BFs, $B_X$, and external
inputs $B_e$, $S_{\rm EW}$, $|V_{ud}|$ and $m_\t$. We have updated
the OPAL $\rho_X(s)$ results, following Ref.~\cite{alphas2}, by first
reconstructing the underlying unit-normalized distributions using
the values for the exclusive-mode BFs and external inputs quoted
by OPAL, and then converting back to $\rho_X(s)$ using the updated
versions of these inputs quoted above.

Contrary to OPAL and ALEPH, Belle already provides the 
unit-normalized number distribution, $(1/N_X)dN_X/ds$, which produces the 
Belle version of $\rho_{2\pi}(s)$ after dividing by the relevant 
external factors and the averaged kinematic weight $w^{\rm av}_T(s;m_\t^2)$
and multiplying by $B_{2\p}$. Again, the Belle data covariance 
matrices are scaled accordingly.

\subsection{\label{algorithm} Data combination algorithm}

Before getting to specific exclusive modes, we describe the underlying
data combination algorithm, which goes back to Ref.~\cite{NNPDF},
and has been used by Keshavarzi, Nomura and Teubner \cite{KNT18} to combine
the world's electroproduction data in their determination of the hadronic
contribution to the inclusive electromagnetic spectral function. We
start by repeating the description of this algorithm given in
Ref.~\cite{alphas20}, after which we discuss the generalization of the
algorithm to include also non-$\chi^2$ fit qualities.

Suppose we wish to combine the spectral data for a particular exclusive mode
from a number of experiments. We begin with choosing a number of
clusters, distributed over the interval $0<s\le m_\t^2$,  assigning a
number of consecutive data points from all experiments to each cluster
$m$, $m=1,\dots,N_{\rm cl}$, with $N_{\rm cl}$ the total number of
clusters. $n_m$ will be the total number of data points in cluster $m$.
Parametrizing the data by pairs $(s_{i},d_{i})$, with $d_{i}$
the data point for the spectral function assigned to the $s$-value
$s_{i}$, we define weighted cluster averages
\begin{equation}
\label{clusters}
s^{(m)}=\sum_{i\in m}\frac{s_{i}}{\s^2_{i}}\Bigg/\sum_{i\in m}
\frac{1}{\s^2_{i}}\ ,
\end{equation}
where the sum is over all data points in cluster $m$ and $\s^2_i$ is the
variance of $d_i$, \ie, the $\s^2_i$ are the diagonal elements of the
covariance matrix $C_{ij}$ for the spectral-function data points $d_i$.
The set of $s^{(m)}$ then constitutes the values of $s$ at which the
combined spectral function $\r^{(m)}$ will be defined.

The values of $\r^{(m)}$ are determined by a linear fit,
minimizing
\begin{equation}
\label{chi2}
\c^2(\r)=\sum_{i=1}^{N}\sum_{j=1}^{N}\left(
d_{i}-R(s_i;\r)\right)C^{-1}_{i j}\left(d_{j}-R(s_j;\r)\right)\ ,
\end{equation}
where $N=\sum_{m=1}^{N_{\rm cl}}n_m$ is the total number of
data points, and the piece-wise linear function $R(s;\r)$ is defined by
\begin{equation}
\label{defR}
R(s;\r)=\r^{(m)}+\frac{s-s^{(m)}}{s^{(m+1)}-s^{(m)}}
\left(\r^{(m+1)}-\r^{(m)}\right)\ ,\quad s^{(m)}\le s\le s^{(m+1)}
\ ,\quad 1\le m<N_{\rm cl}\ ,
\end{equation}
where $\r$ is the vector of fit parameters $\r^{(m)}$, $m=1,\dots,N_{\rm cl}$.
At the boundaries, we extrapolate:
\begin{eqnarray}
\label{defR2}
    R(s;\r)&=&\r^{(N_{\rm cl}-1)}+\frac{s-s^{(N_{\rm cl}-1)}}
{s^{(N_{\rm cl})}-s^{(N_{\rm cl}-1)}}
\left(\r^{(N_{\rm cl})}-\r^{(N_{\rm cl}-1)}\right)
\ ,\quad s\ge s^{(N_{\rm cl})}\ ,
\\
R(s;\r)&=&
\r^{(1)}+\frac{s-s^{(1)}}{s^{(2)}-s^{(1)}}\left(\r^{(2)}-\r^{(1)}\right)
\ ,\qquad\qquad\qquad\qquad\ s\le s^{(1)}\ .\nonumber
\end{eqnarray}
Writing $R$ as a rectangular matrix $R_{im}$ such that
\begin{equation}
\label{Rmatrix}
\sum_{m=1}^{N_{\rm cl}}R_{im}\r^{(m)}=R(s_i;\r)
\end{equation}
and using vector notation, the minimum of Eq.~(\ref{chi2}) is obtained at
\begin{eqnarray}
\label{minchi2}
\vec\r&=&M\vec{d}\ ,\\
M&=&(R^TC^{-1}R)^{-1}(R^TC^{-1})\ ,\nonumber
\end{eqnarray}
while the covariance matrix, $\cc_{mn}$, of the cluster $\vec\r$ parameters, is given by
\begin{equation}
\label{clcovM}
\cc=MCM^T\ .
\end{equation}
Using Eq.~(\ref{minchi2}), this reduces to
\begin{equation}
\label{clcov}
\cc^{-1}=R^TC^{-1}R\ .
\end{equation}
Note that the length of the parameter vector $\vec\r$ is smaller than
that of the data vector~$\vec{d}$.

The algorithm described above applies if the full data covariance matrix
is available and positive definite (and thus invertible). In our
application to the data, this is always the case for the $2\p$
data combination; the $\p^-\p^0$ ALEPH, OPAL and Belle covariance
matrices are all positive definite, and the corresponding correlation
matrices do not appear to have ``suspiciously'' small eigenvalues.
However, this is not the case for the two $4\p$ modes (as already
observed in Ref.~\cite{alphas20}), and this leads us to consider the
minimization of a fit quality
\begin{equation}
\label{defQ2}
Q^2=(\vec{d}-R\vec\r)^TW^2(\vec{d}-R\vec\r)\ ,
\end{equation}
with the positive definite matrix $W^2$ replacing $C^{-1}$ in Eq.~(\ref{chi2}).
In such cases we use the methods of Ref.~\cite{BS} to obtain the parameter
covariance matrix $\cc$ and compute a $p$-value for the fit. 
In this case, one finds 
\begin{equation}
\label{W2fit}
M=(R^TW^2R)^{-1} (R^TW^2)\ ,
\end{equation}
while $\cc$ is still obtained from Eq.~(\ref{clcovM}) with $M$ now given by Eq.~(\ref{W2fit}).
Of course, for $W^2=C^{-1}$, one recovers Eqs.~(\ref{chi2}) and~(\ref{clcov}).

As in Ref.~\cite{alphas20}, in order to assess the local quality of our fits, we
also, for each fit, compute the $\c^2$ (or $Q^2$) value and 
corresponding $p$-value for each cluster in the fit, \ie, Eq.~(\ref{chi2}) 
restricted to each of the clusters labeled by $m$ for all 
$m=1,\dots,N_{\rm cl}$, to check that there are no large local 
discrepancies between the different data sets.
Such local $\c^2$ (or $Q^2$) computations will always use the full corresponding
block-diagonal data covariance matrices, where the blocks are defined
by the clusters.\footnote{In the case of the $\p^-3\p^0$ mode, only the
diagonal part of the OPAL covariance matrix is used, see Sec.~\ref{4picomb} below.}
Note however, that in the case of a fully correlated
global fit, the full $\c^2$~(\ref{chi2}) is not equal to the sum of the
local $\c^2$ values, as in general there will be non-vanishing correlations
between data in different clusters.

For the $2\p$ combination, described in Sec.~\ref{2picomb} below, we will
combine the ALEPH, OPAL and Belle unit-normalized data, divided
by $w_T(s;m_\t^2)$, before multiplying with the factor
$B_{\p^-\p^0}/\cf$ required to produce the $2\pi$ spectral function.
In the case of the $4\p$ combination described in Sec.~\ref{4picomb}, 
however, we will need to first form the ALEPH and OPAL
versions of the sum of $2\p^-\p^+\p^0$ and $\p^-3\p^0$ spectral
functions, including the factors $B_{2\p^-\p^+\p^0}/\cf$ and
$B_{\p^-3\p^0}/\cf$ in the process, before combining the ALEPH and
OPAL results. In this case, the errors on the two BFs lead to a
potential bias, known as the d'Agostini bias Ref.~\cite{Abias}. We will
explain how this issue is handled in Sec.~\ref{4picomb} below.

\subsection{\label{2picomb} \begin{boldmath} ALEPH, OPAL and Belle $2\p$ data
combination\end{boldmath}}
For the $\pi^{-}\pi^{0}$ exclusive mode, the ALEPH \cite{ALEPH13},
OPAL \cite{OPAL} and Belle \cite{Belle} unit-normalized number
distribution data sets have positive definite covariance matrices, which
allows us to combine them using the standard $\chi^2$ fit analysis,
described in Sec.~\ref{algorithm}. There are two advantages to performing
the combination of the unit-normalized distributions first, rather than
directly combining the three $2\pi$ spectral functions. First,
this ensures the combination is not subject to a d'Agostini bias
\cite{Abias}, which could arise from the correlations induced by the overall normalization 
with the BF. 
Second, since no BF needs to be included in the data before performing
the combination, the effect of future changes in $B_{\p^-\p^0}$ can
be incorporated by a global rescaling, without having to re-run the
entire combination procedure.

\begin{figure}[t]
\begin{center}
\includegraphics*[width=15.25cm]{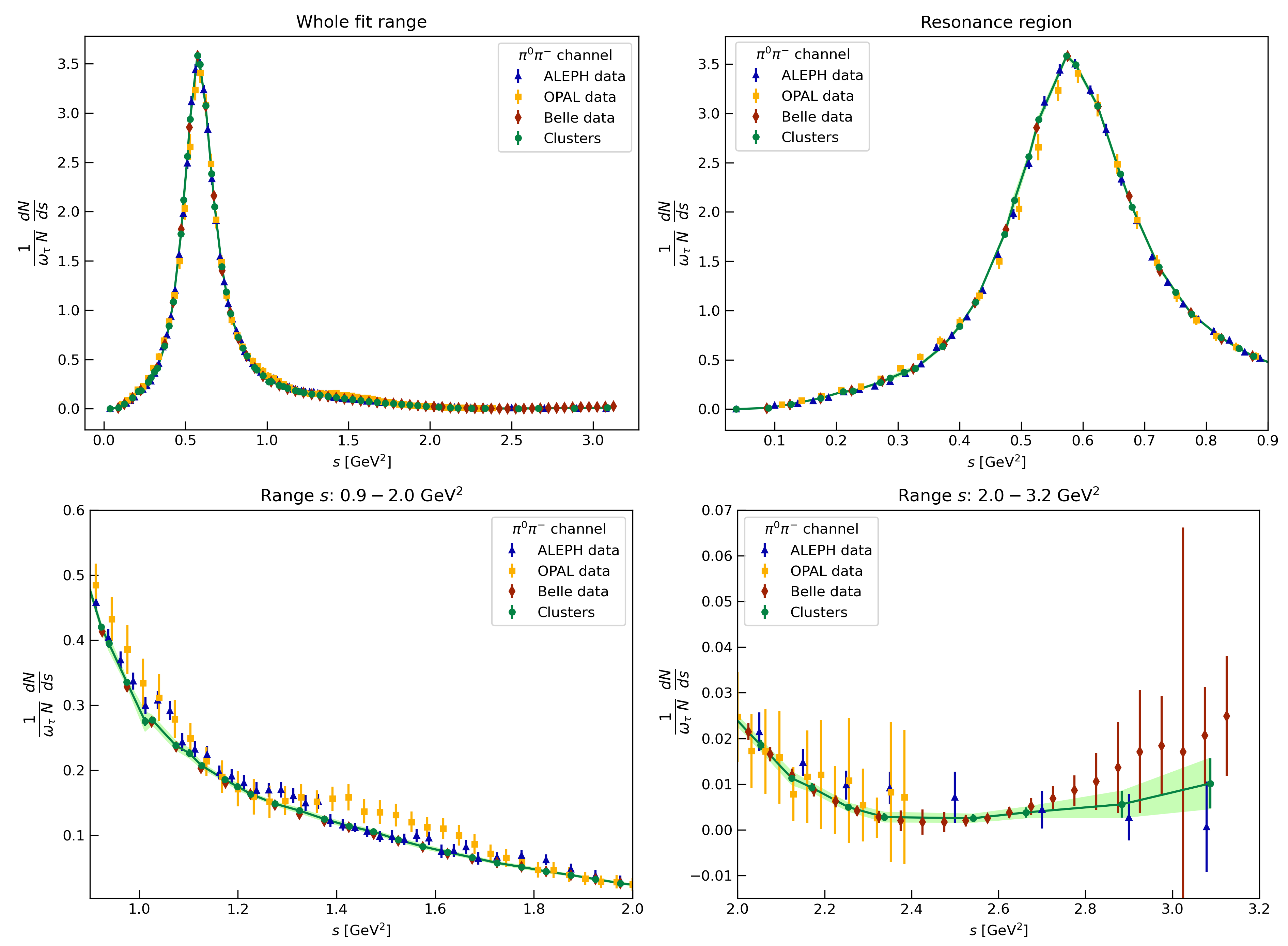}
\end{center}
\begin{quotation}
\floatcaption{2picomb-fig}%
{{\it Results of the fit for the 63-cluster combination of the $\p^-\p^0$ 
exclusive-mode unit-normalized distribution divided by
the $s$-dependent kinematic weight factor. The error bars represent
the uninflated errors, while inflated errors are represented
by the green band.}}
\end{quotation}
\vspace*{-4ex}
\end{figure}

\begin{figure}[t!]
\vspace*{4ex}
\begin{center}
\includegraphics*[width=8.25cm]{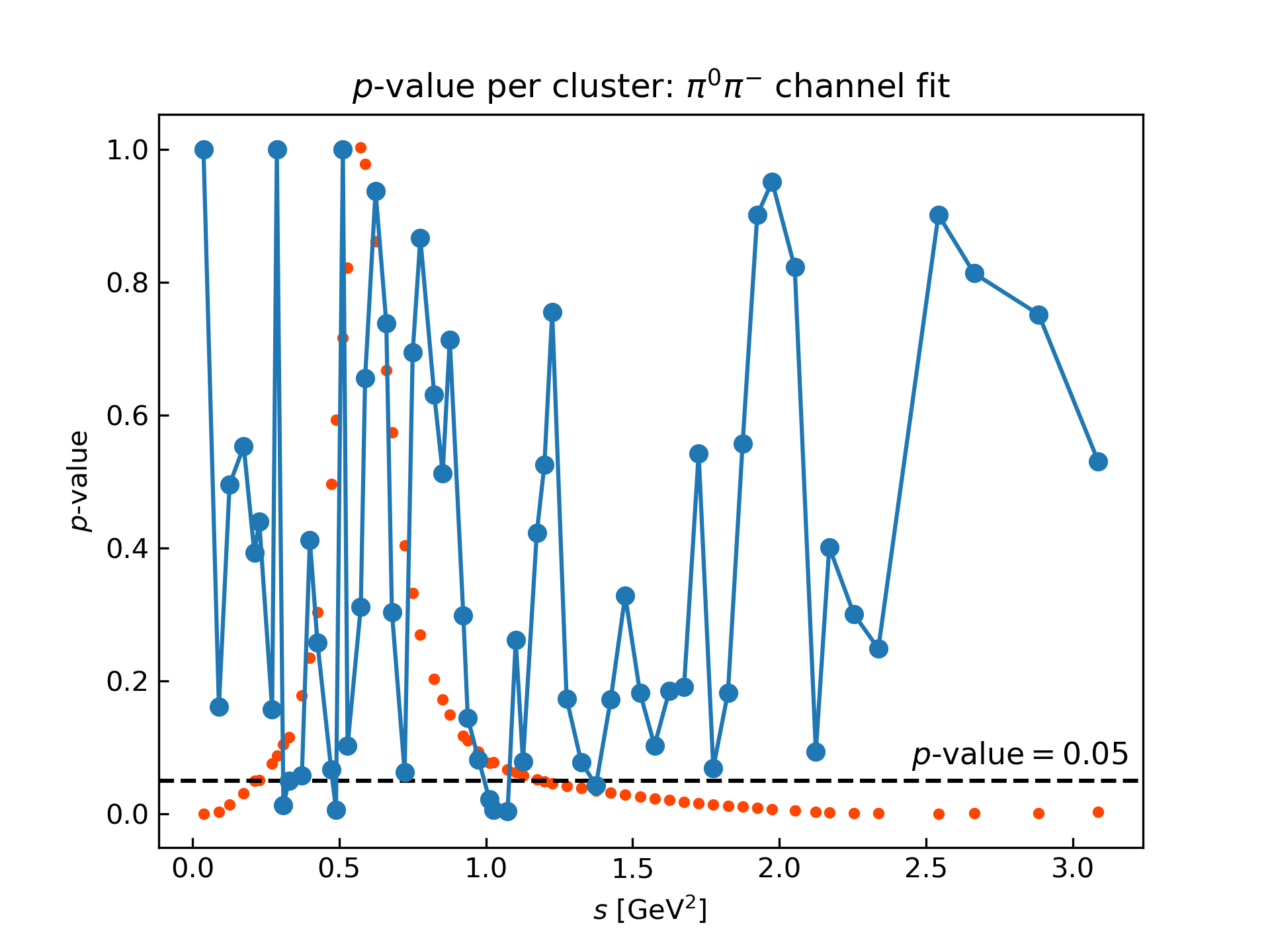}
\end{center}
\begin{quotation}
\floatcaption{2pipvalue-fig}%
{{\it Local $p$-value for each cluster, for the 63-cluster
$\pi^{-}\pi^{0}$ unit-normalized combined spectral function. 
The red dots show, up to an
arbitrary overall normalization,
the corresponding combined 
$2\pi$ spectral function.}}
\end{quotation}
\vspace*{-4ex}
\end{figure}

An essential step in the combination method described in Sec.~\ref{algorithm}
is choosing a reasonable strategy for dividing the data into
clusters. Since the Belle data set is considerably more precise
than those of ALEPH and OPAL, a first requirement we impose is that each 
cluster contain at least one Belle data point. Due to the
rapid variation of the spectral function around the $\rho$ peak, a
finer grid of clusters is needed in this region, while clusters with
more data points are used in the higher-$s$ region where the data
are sparser, more slowly varying, and have larger errors. With this
strategy, from a comparison of the experimental uncertainties
of the three data sets, we expect Belle to dominate the
combination.

Our combined $2\pi$ results are obtained from a configuration
with 63 clusters, which produces a global fit quality of
\begin{equation}
\label{2pi fit quality}
\chi^2/\textrm{dof} = 168.65/150\ , \quad p\textrm{-value} = 14\%\ .
\end{equation}
The results of this combined fit are shown by the green points in
Fig.~\ref{2picomb-fig}, together with the corresponding single-experiment 
ALEPH, OPAL and Belle inputs. We have considered alternate
choices of clusters following the same principles, and always obtained 
very similar results.

We have also computed local $\c_{\rm cl}^2$ values, \ie, $\c^2$ values
for each cluster, from our fit results.  The goal of the local $\c_{\rm cl}^2$
calculation is to check whether the three different data sets
agree also locally, in each cluster. With the number of degree
of freedom in each cluster known, we can convert the local
$\c_{\rm cl}^2$ values into local $p_{\rm cl}$-values, which are shown
in Fig.~\ref{2pipvalue-fig}.\footnote{There are three clusters with only
one data point, \ie, with no degrees of freedom. For such clusters, we set
the $p_{\rm cl}$-value equal to 1.} The $p_{\rm cl}$-values of the
clusters are generally good, with the exception of three clusters for which
$p_{\rm cl}\sim 0.005$. We conclude that the three
data sets are statistically compatible. For clusters with a local
$\c_{\rm cl}^2$ values larger than 1, we account for the tension between
the data sets in each such cluster by inflating the errors,
multiplying them by $\sqrt{\c_{\rm cl}^2}$. The resulting inflated
errors are shown by the green bands in Fig.~\ref{2picomb-fig}.
No error rescaling is performed for clusters with local $\c^2$
values smaller than 1. As can be seen in Fig.~\ref{2picomb-fig},
the difference between the error bars on the green points
and the green band is barely visible.

Figure~\ref{2piratio-fig} shows the ratio of each of the three
single-experiment $2\pi$ spectral functions to the combined
version, in the region around the $\r$ peak. The impact of the
Belle data is obvious: they dominate the final combined spectrum,
especially in the $\rho$-peak region. While the quality of the fit is good, 
it can be seen that in the region just beyond the $\r$ peak, the 
Belle data lie systematically somewhat below the combined 
$\p^-\p^0$ distribution. Figure~\ref{2piratio-fig} can be compared to 
similar figures in Ref.~\cite{ALEPH13}.

\begin{figure}[t]
\vspace*{4ex}
\begin{center}
\includegraphics*[width=15.25cm]{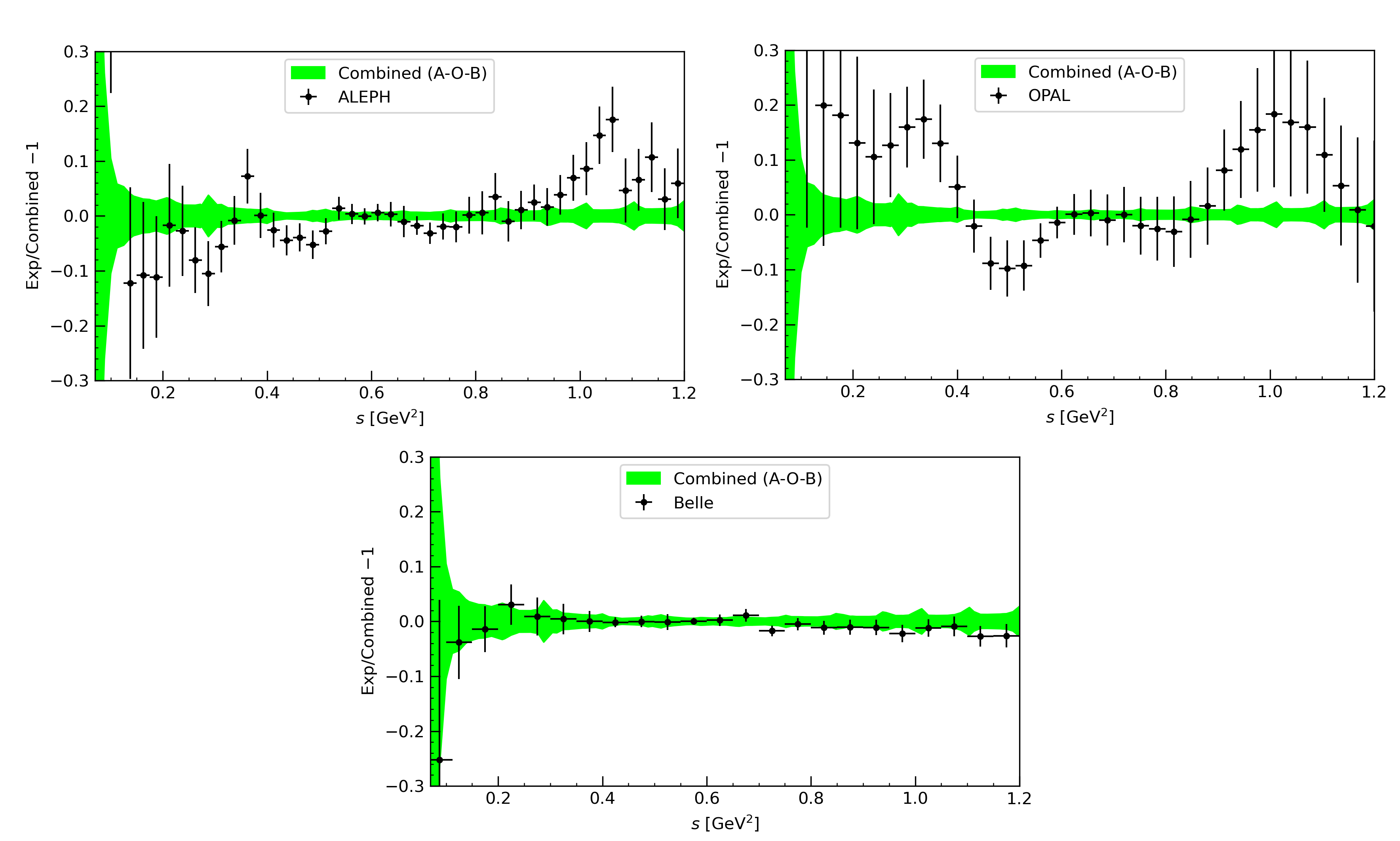}
\end{center}
\begin{quotation}
\floatcaption{2piratio-fig}%
{{\it Ratio between experimental $2\pi$ data and the associated
$2\pi$ combination, in the $\rho$ peak region.}}
\end{quotation}
\vspace*{-4ex}
\end{figure}

\subsection{\label{4picomb} \begin{boldmath} ALEPH and OPAL $4\p$ data 
combination\end{boldmath}}
In order to prepare the ALEPH \cite{ALEPH13} and OPAL \cite{OPAL} data
sets, we discard the small number of $2\pi^{-}\pi^{+}\pi^{0}$ and
$\pi^{-}3\pi^{0}$ points having total or systematic error equal
to zero, assuming such points cannot be interpreted as reliably
determined data.

\begin{figure}[t]
\vspace*{4ex}
\begin{center}
\includegraphics*[width=10.25cm]{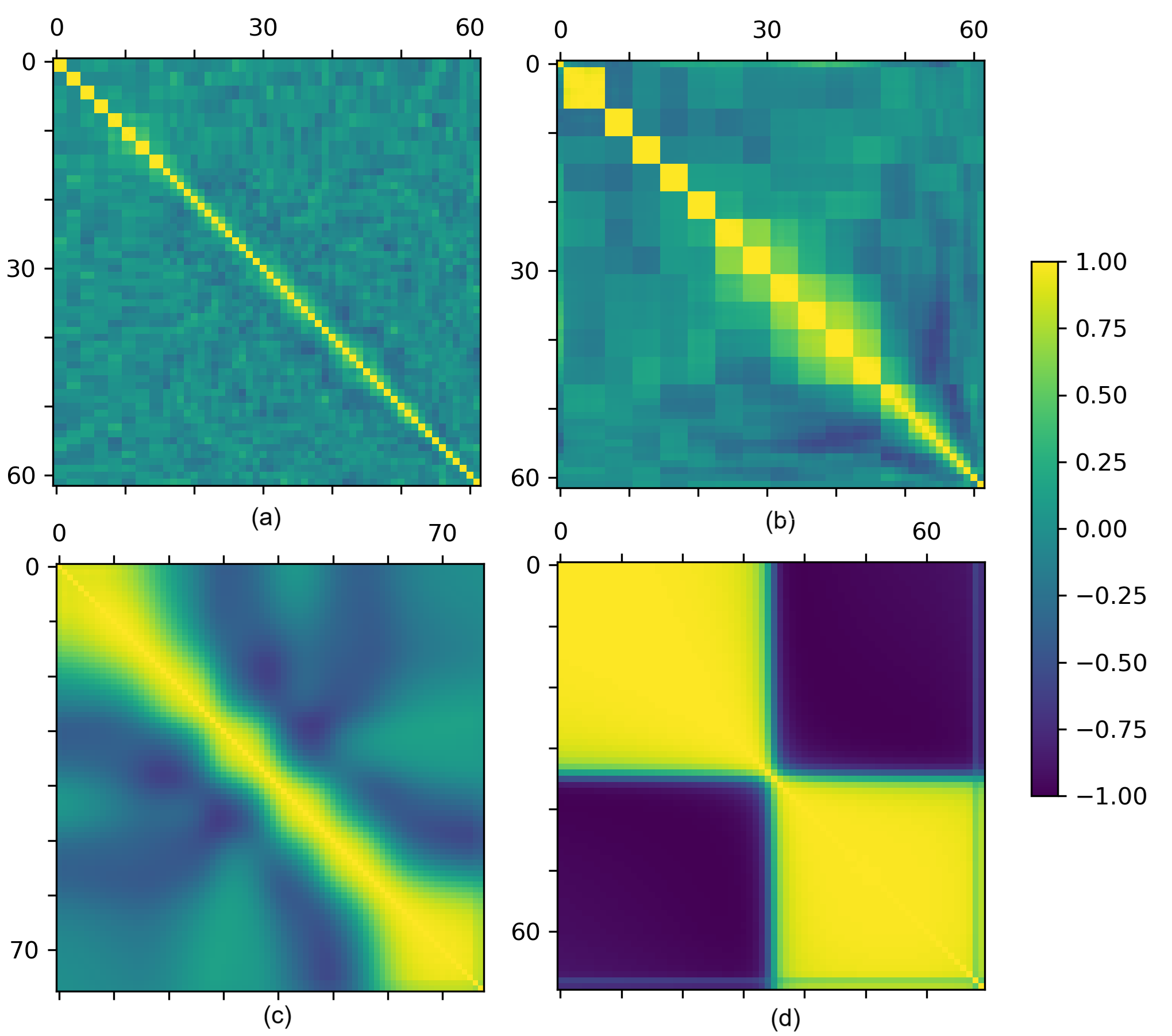}
\end{center}
\begin{quotation}
\floatcaption{Correlations-fig}%
{{\it Correlations of the unit-normalized, $4\pi$ exclusive-mode
number distribution data sets. (a) ALEPH $ 2\pi^{-}\pi^{+}\pi^{0}$,
(b) ALEPH $\pi^{-}3\pi^{0}$, (c) OPAL $ 2\pi^{-}\pi^{+}\pi^{0}$ and
(d) OPAL $\pi^{-}3\pi^{0}$.}}
\end{quotation}
\vspace*{-4ex}
\end{figure}

Unlike the case of the $2\pi$ data, the unit-normalized
$2\pi^{-}\pi^{+}\pi^{0}$ and $\pi^{-}3\pi^{0}$ covariance matrices
of ALEPH and OPAL are ill-behaved. The ALEPH
$2\pi^{-}\pi^{+}\pi^{0}$ and $\pi^{-}3\pi^{0}$ correlation matrices
have 1 and 14  negative eigenvalues of order $10^{-9}$ and $10^{-6}$ or less
in magnitude, respectively, while the OPAL
$\pi^{-}3\pi^{0}$ covariance matrix has 1 negative eigenvalue with magnitude of order $10^{-2}$.  

Because of these issues with the exclusive-mode $4\pi$ covariance
matrices, combining their unit-normalized number distributions
following the same procedure as employed for the $\pi^{-}\pi^{0}$ case
(using fully correlated fits) is not possible. However, summing the
$\pi^{-}3\pi^{0}$ and $2\pi^{-}\pi^{+}\pi^{0}$ contributions
(properly normalized with the corresponding BFs) to produce
the total $4\pi$ spectral function for each of ALEPH and OPAL,
we find that both of the associated total $4\p$ covariance
matrices are positive definite. The total $4\p$ covariance matrix
for OPAL, however, has very small (positive) eigenvalues indicating
very strong correlations, and a standard $\chi^2$ fit employing
Eq.~(\ref{chi2}) leads to a global ALEPH plus OPAL combination
with a $p$-value less than $0.1\%$.

We show the four exclusive-mode $4\p$ data correlation matrices in
Fig.~\ref{Correlations-fig}. As is immediately evident, the
qualitative pattern is very different in the OPAL $\pi^{-}3\pi^{0}$
case. It is this $\pi^{-}3\pi^{0}$ covariance matrix that
leads to the poor combined fit.\footnote{This poor behavior is most
likely due to the difficulty of identifying neutral pions experimentally,
leading to strong correlations with the $\p^-\p^0$ and
$\p^-2\p^0$ exclusive-mode results. We thank Sven Menke for a detailed discussion
of this point and the underlying experimental issues.}

We thus depart from a strict $\c^2$ fit and 
turn to a fit using Eq.~(\ref{defQ2}) with the ALEPH and OPAL
exclusive-mode $4\pi$ spectral functions as input.
The $W^2$ matrix used in Eq.~(\ref{defQ2}) is constructed as follows. We
replace the OPAL $\pi^{-}3\pi^{0}$ covariance matrix
with a diagonal truncation, thus removing
the strong, but poorly determined, correlations.
The matrix $W^2$ of Eq.~(\ref{defQ2}) is then constructed from the unmodified ALEPH 
$2\pi^- \pi^+\pi^0$ and $\pi^- 3\pi^0$
covariance matrices, the unmodified OPAL $2\pi^{-}\pi^{+}\pi^{0}$ covariance matrix, and the
diagonally truncated OPAL $\pi^{-}3\pi^{0}$ covariance matrix.
As already mentioned, we keep the full,
original data covariance matrices for error propagation, \ie, the
$\vec{\r}$ parameter covariance matrix $\cc$ is computed 
with the full data covariance matrix $C$,
using Eq.~(\ref{clcovM}) with $M$ as given in Eq.~(\ref{W2fit}).

Since the $2\pi^{-}\pi^{+}\pi^{0}$ and $\pi^{-}3\pi^{0}$ spectral
functions which serve as inputs to the fit have global normalizations set
by the corresponding exclusive-mode BFs, our fit procedure potentially
suffers from the d'Agostini bias. In principle, one would like to use the
iterative procedure proposed in Ref.~\cite{NNPDF} in order to correct for this
bias, but, since after the first iteration only the sum of the two
exclusive-mode $4\p$ contributions is available, we have to resort
to an approximate method, which is described in App.~\ref{Abiascorr}.
Our results presented below have been obtained using this approximate
bias-correction method. The fact that the correction shifts
the uncorrected result by only a small amount gives us confidence
that the bias is not a significant effect.

\begin{figure}[t]
\vspace*{4ex}
\begin{center}
\includegraphics*[width=15.25cm]{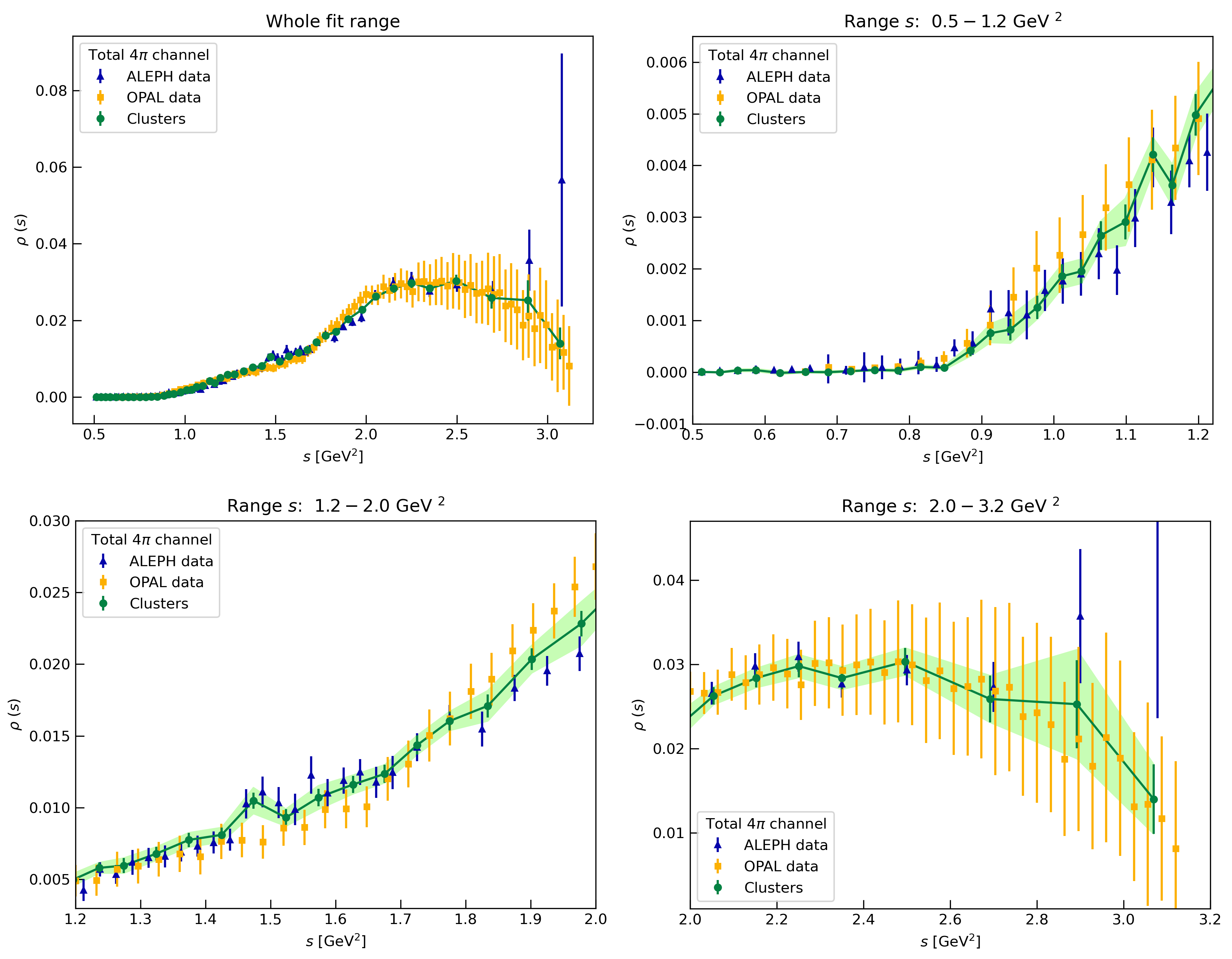}
\end{center}
\begin{quotation}
\floatcaption{4picomb-fig}%
{{\it Results (green points) of the fit for the 46-cluster combination
of the ALEPH and OPAL two-mode $4\pi$ spectral function results. The
error bars on the green points represent the uninflated errors, while
inflated errors are represented by the green band.}}
\end{quotation}
\vspace*{-4ex}
\end{figure}

Our result for the ALEPH plus OPAL two-mode $4\pi$ spectral
function combination has 46 clusters, and a global fit quality of
\begin{equation}
\label{4pi fit quality}
p\textrm{-value} = 10.3\%\ ,
\end{equation}
computed following the procedure described in Ref.~\cite{BS}. The resulting
$4\pi$ combination is shown in Fig.~\ref{4picomb-fig}, and the local
$p_{\rm cl}$-values, calculated using the methods of Ref.~\cite{BS},
in Fig.~\ref{4pipvalue-fig}. As in the case of the $2\pi$ fit,
there are no clusters with unacceptably small local $p_{\rm cl}$-values
(with none having a $p_{\rm cl}$-value~$<0.01$).
As before, green bands in the figure show inflated error bars computed
as in Sec.~\ref{2picomb}; differences with the uninflated errors
represented by the green error bars are barely visible.

\begin{figure}[t]
\vspace*{4ex}
\begin{center}
\includegraphics*[width=9.25cm]{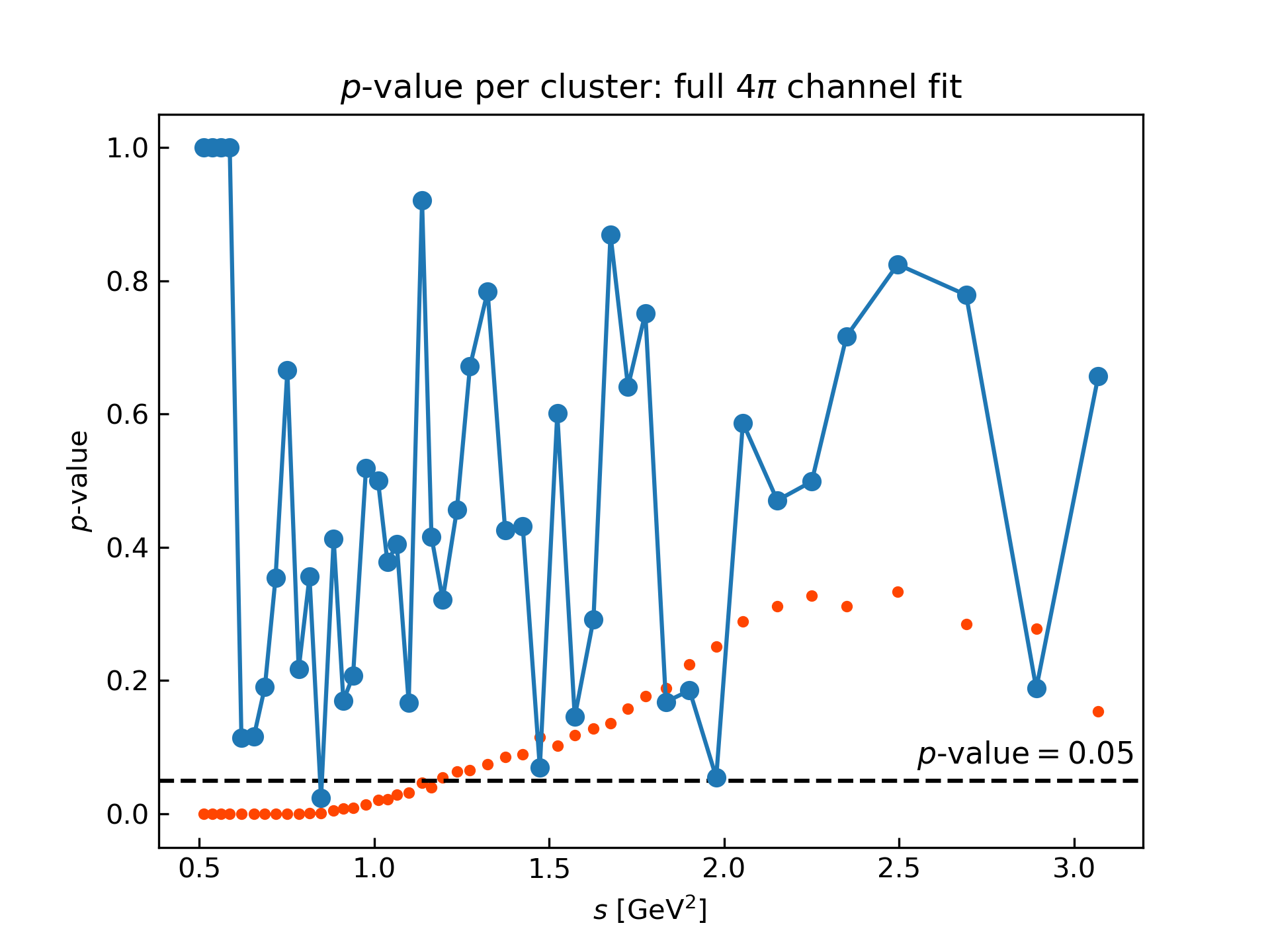}
\end{center}
\begin{quotation}
\floatcaption{4pipvalue-fig}%
{{\it Local $p$-value for each cluster, for the 46-cluster $4\pi$
spectral function combination.  The red dots show, up to an
arbitrary overall normalization,
the corresponding combined 
$4\pi$ spectral function.}}
\end{quotation}
\vspace*{-4ex}
\end{figure}

\subsection{\label{full} \begin{boldmath} Full $2\pi$ and $4\p$ data
combination\end{boldmath}}
In this section, we outline how we combine the $\p^-\p^0$ spectral
function obtained in Sec.~\ref{2picomb} with the total $4\p$ spectral function
obtained in Sec.~\ref{4picomb}. This would be trivial if there were no
correlations between the $2\p$ and $4\pi$ results: one would simply
construct the combined $2\pi$ spectral function from the chosen input
external constants and unit-normalized $2\p$ result of Sec.~\ref{2picomb},
interpolate the $4\p$ spectral function of Sec.~\ref{4picomb} to the $s$
values of the $2\p$ spectral-function clusters, and add the two results.

There are, however, two sources of correlation between the $2\p$
and $4\p$ spectral functions. First, there are the correlations inherent
to the ALEPH and OPAL experiments themselves; see, for instance, the
discussion of the OPAL $\p^-3\p^0$ covariance matrix in Sec.~\ref{4picomb}.
These correlations have not been provided by the experiments, and we thus
have to assume that they are relatively small.\footnote{This assumption,
though not made explicit, was also made in Ref.~\cite{alphas20}.} Second,
there are correlations induced by the correlations between the different
exclusive-mode BFs, {\it cf.} the $r$-values given in Eq.~(\ref{BX}).
These correlations are also quite small, but since they have been
provided by HFLAV \cite{hflav22}, there is no reason not to take them
into account. 

Appendix~\ref{2pi4pi} provides a detailed description of our procedure for 
including them and generating the $2\p +4\p$ spectral function, 
$\rho_{2\p + 4\p}(s)$, and associated covariance matrix. 
If information on other sources of correlation between the $2\p$ and $4\p$
data were to become available, those correlations can be taken into account
following the same strategy.

With the $2\p +4\p$ combination complete, the last step remaining in the construction of the inclusive
spectral function is the addition of residual-mode contributions. This step is discussed in Sec.~\ref{specfun}. 

\subsection{\label{specfun} \begin{boldmath} The inclusive non-strange $V$ spectral function\end{boldmath}}
With the $2\pi+4\pi$ spectral function, generated using the algorithm
described in App.~\ref{2pi4pi}, in hand, the next step is to add the remaining,
residual-mode contributions to produce the total inclusive spectral
function. In the original ALEPH and OPAL publications, lacking reliable
experimental determinations, Monte-Carlo input was used in
obtaining the contributions from 
these modes. Our approach to these contributions was
already discussed in Ref.~\cite{alphas20}, and details of how the exclusive
residual-mode contributions were obtained from (i) 2019 BaBar
$\t\to K\bar{K}\n_\t$ data, and (ii) exclusive-mode $e^{+}e^{-}\rightarrow \textrm{hadrons}$
cross-sections using CVC, can be found in Ref.~\cite{alphas20}.  Since the work
of Ref.~\cite{alphas20}, there have been no significant changes in the
relevant experimental branching fractions, and no new data sets
that would impact our previous work. We thus employ for the
residual-mode contributions the results of Ref.~\cite{alphas20}, re-interpolated
onto the grid of $s$ values of the new 63-cluster set.

\begin{figure}[t!]
\vspace*{4ex}
\begin{center}
\includegraphics*[width=13cm]{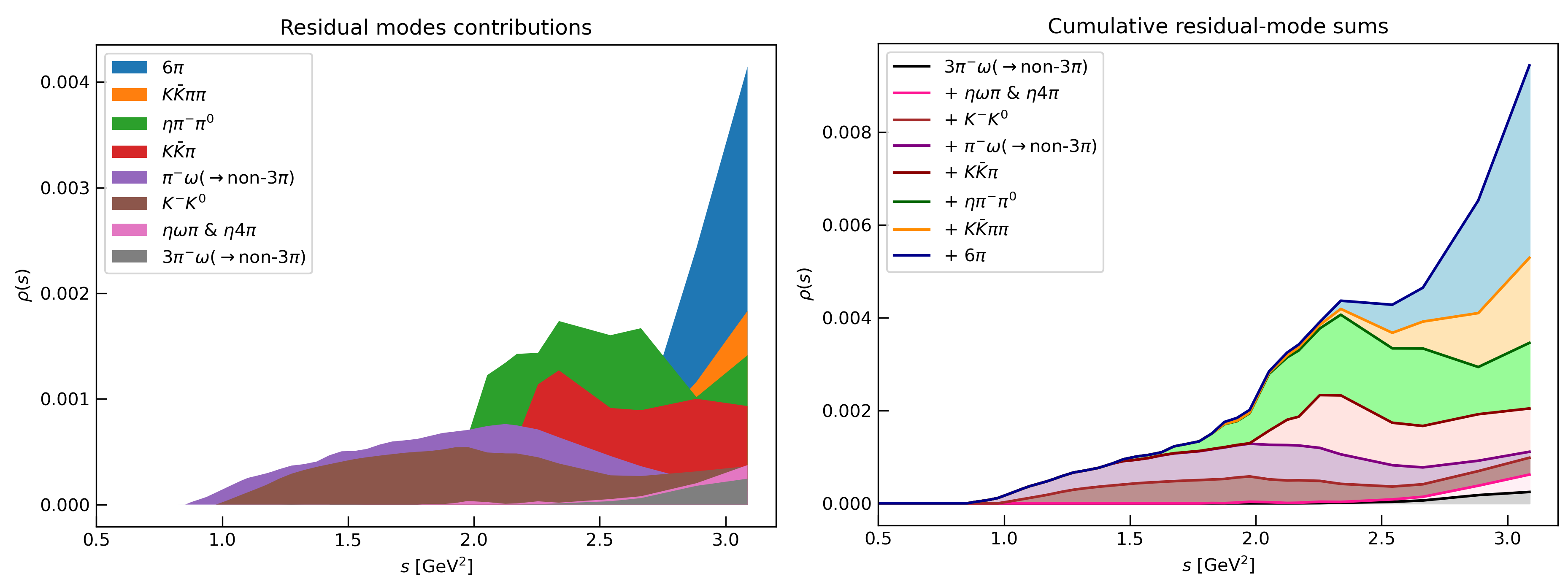}
\end{center}
\begin{quotation}
\floatcaption{resmodes-fig}%
{{\it Residual-mode contributions to the inclusive spectral function.
The left panel shows individual residual-mode contributions, the right
panel the corresponding cumulative residual-mode totals.}}
\end{quotation}
\vspace*{-4ex}
\end{figure}

The residual modes for which spectral function contributions can be
obtained, using CVC, from exclusive-mode electroproduction cross-sections
are $\p^- \o (\rightarrow \mbox{non-}3\p)$ \cite{sndomegapi16,babaromegapi17},
$\eta \p^-\p^0$ \cite{Belleetapipi09,sndetapipi1518,sndetapipi15182,babaretapipi18,cmd3etapipi19,cmd3etapipicovs},
$K\bar{K}\p$ \cite{babarepemkkbarpi},
$6\p$ \cite{babarsigma6pi06,thankssolodov,cmd3sigma3pim3pip13,cmd3sigmaeta3pi17,sndsigmapimpip4pi019,
sndsigmaeta3pi19},
$K\bar{K}\p\p$ \cite{babarepemkkbarpi},
$(3\p)^-\o (\rightarrow {\rm \mbox{non-}3}\p)$ \cite{babarsigma6pi06},
$\p^-\eta\o(\rightarrow {\rm \mbox{non-}3}\p)$ \cite{babarepemetapimpip2pi018} and
$\eta 4\p$ \cite{babarepemetapimpip2pi018,babarepemeta2pim2pip07}. The
remaining residual-mode contribution, that of $K^- K^0$, is, as noted above,
obtained directly from the $\t$ decay results of Ref.~\cite{babarkkbartau18}.
The contribution of each residual mode to the residual-mode spectral
function total is shown in Fig.~\ref{resmodes-fig}. In order to add
these contributions to the combined $2\pi+4\pi$-mode sum, we
have linearly interpolated each residual mode data set to the
central values $s^{(m)}$ of the 63 $2\pi$ mode clusters. 
The residual-mode covariance matrices were interpolated
accordingly.

The final result for the inclusive isovector $V$ spectral function,
$\rho_{ud; V}(s)$, is given by the sum of the $\rho_{2\pi+4\pi}$
contributions from Secs.~\ref{2picomb}, \ref{4picomb} and \ref{full} and
the interpolated residual-mode results. In the left panel of
Fig.~\ref{finalspec-fig} we show the individual contributions
($2\pi$, $4\pi$ and residual mode total) to $\rho_{ud; V}(s)$,
and in the right panel, the total inclusive spectral function
(with the light-blue band representing the inflated errors).
Numerical versions of these result, with both uninflated and
inflated errors, are presented in Tab.~\ref{tabspec}. 
The accompanying covariance matrix is too large to be
tabulated here, but can be requested from the authors.

\begin{table}[t]
{\footnotesize
\begin{center}
\begin{tabular}{|c|c||c|c||c|c||c|c|}
\hline
$s$  & $2\pi^2\rho_{ud;V}(s)$ & $s$  & $2\pi^2\rho_{ud;V}(s)$
& $s$  & $2\pi^2\rho_{ud;V}(s)$ & $s$  & $2\pi^2\rho_{ud;V}(s)$ \\
\hline
0.038 & 0.000(00)(00) & 0.529 & 2.286(20)(28) & 1.026 & 0.258(05)(07) & 1.725 & 0.353(13)(13)\\ 
0.090 & 0.009(02)(03) & 0.574 & 2.790(16)(16) & 1.073 & 0.245(05)(07) & 1.775 & 0.383(13)(13)\\ 
0.126 & 0.038(02)(02) & 0.588 & 2.720(19)(19) & 1.102 & 0.243(08)(10) & 1.825 & 0.399(15)(19)\\ 
0.175 & 0.086(03)(03) & 0.624 & 2.395(15)(15) & 1.126 & 0.245(06)(07) & 1.875 & 0.441(14)(17)\\
0.211 & 0.138(05)(05) & 0.661 & 1.857(17)(17) & 1.173 & 0.233(07)(07) & 1.925 & 0.478(14)(20)\\ 
0.227 & 0.142(04)(04) & 0.680 & 1.597(12)(13) & 1.199 & 0.246(08)(09) & 1.975 & 0.509(17)(29)\\  
0.271 & 0.212(05)(06) & 0.723 & 1.123(08)(12) & 1.226 & 0.249(07)(07) & 2.053 & 0.589(21)(21)\\ 
0.288 & 0.244(09)(09) & 0.750 & 0.925(10)(10) & 1.275 & 0.247(10)(10) & 2.125 & 0.621(20)(20)\\
0.310 & 0.291(07)(17) & 0.777 & 0.751(07)(07) & 1.325 & 0.256(09)(10) & 2.170 & 0.640(22)(22)\\ 
0.329 & 0.322(05)(09) & 0.824 & 0.566(05)(05) & 1.375 & 0.266(10)(10) & 2.254 & 0.668(27)(27)\\ 
0.373 & 0.496(06)(10) & 0.853 & 0.482(08)(08) & 1.425 & 0.266(10)(10) & 2.337 & 0.652(25)(25)\\ 
0.400 & 0.654(09)(09) & 0.877 & 0.424(04)(04) & 1.475 & 0.308(11)(18) & 2.542 & 0.664(32)(32)\\
0.426 & 0.846(06)(07) & 0.923 & 0.344(04)(05) & 1.525 & 0.277(12)(12) & 2.664 & 0.618(50)(50)\\ 
0.473 & 1.380(10)(14) & 0.938 & 0.325(07)(09) & 1.575 & 0.297(12)(16) & 2.884 & 0.632(102)(125)\\ 
0.489 & 1.649(17)(44) & 0.975 & 0.288(05)(06) & 1.625 & 0.308(12)(13) & 3.087 & 0.448(93)(93)\\ 
0.513 & 1.994(23)(23) & 1.012 & 0.255(07)(12) & 1.675 & 0.319(12)(13) & & \\
 \hline
\end{tabular}
\end{center}
\floatcaption{tabspec}{{\it Total inclusive spectral function multiplied
by $2\p^2$. The first and second errors are, respectively, the
uninflated and inflated ones. For reference, the parton-model value
for $2\pi^2\rho_{ud;V}(s)$ is $\half$.}}
}\end{table}

\begin{figure}[t]
\vspace*{4ex}
\begin{center}
\includegraphics*[width=15.25cm]{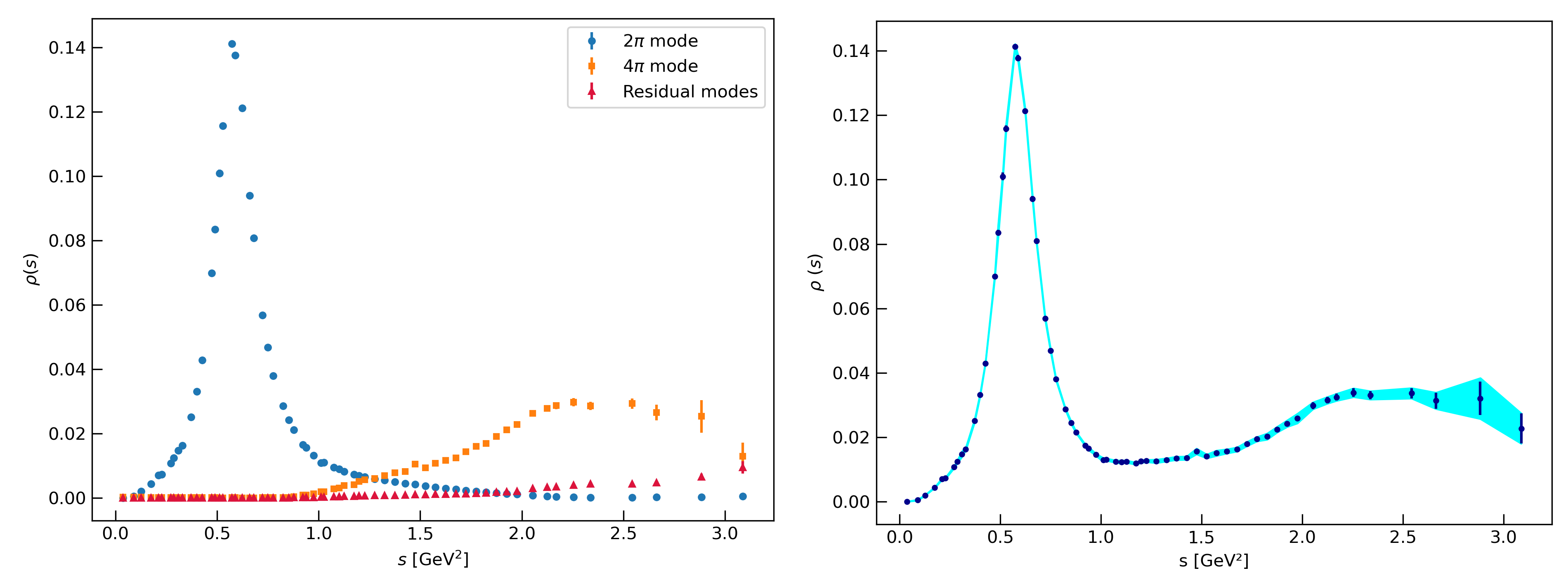}
\end{center}
\begin{quotation}
\floatcaption{finalspec-fig}%
{{\it The non-strange vector-isovector spectral function,
$\rho_{ud; V}(s)$. Left panel: separate $2\pi$, $4\pi$ and
residual-mode contributions. Right panel: the total inclusive
result. The light-blue band represents the inflated errors.}}
\end{quotation}
\vspace*{-4ex}
\end{figure}

\section{\label{strong coupling} The strong coupling}
In this section, we employ our new inclusive, non-strange $V$ spectral
function to evaluate the spectral moments $I_{\rm exp}^{(w)}(s_0)$
of Eq.~(\ref{cauchy}), with the weights defined in Eq.~(\ref{weights}), in order
to determine $\a_s(m_\t^2)$. Our analysis follows the same strategy as
in Ref.~\cite{alphas20}. We briefly comment on our strategy
in Sec.~\ref{strategy}, and present our results in Sec.~\ref{results}.
Section~\ref{analysis} provides further analysis and discussion.

\subsection{\label{strategy} Strategy}
In the first, and main step of our analysis, we use the
spectral integrals, $I_{\rm exp}^{(w_0)}(s_0)$, of the
$w_0=1$ FESR to determine $\a_s(m_\t^2)$ and the DV parameters
$\d$, $\g$, $\a$ and $\b$, considering all $s_0$ in the
interval between a value $s_0^{\rm min}$ and the maximum value
available for our 63-cluster spectral function. We
vary $s_0^{\rm min}$, and look for a region in which the fits
(as measured by the $p$-values) are good, at the same time demanding
stability of the results with respect to $s_0^{\rm min}$. We then
average $\a_s(m_\t^2)$ over the fits in this stability region.
While the spectral integrals for nearby values of $s_0$ are highly
correlated, there is no problem with carrying out these fits: we
find a ``plateau'' region of $s_0^{\rm min}$ values with stable,
high-quality fits. The central value of our final result, as well
as the statistical error, are obtained from these fits, taking
into account the correlations between the results for different
$s_0^{\rm min}$.

In the second step, we perform combined fits using two different weights,
one of which is always the weight $w_0$, and the other is $w_i$, with
$i=2,3,4$. As the weights $w_{2,3,4}$ are pinched, the corresponding
spectral integrals are less sensitive to the DV parameters, making
it more difficult to reliably determine these parameters in fits
that do not also include $I_{\rm exp}^{(w_0)}(s_0)$. The motivation
for performing such two-weight, multiple-$s_0$ fits is that the addition
of the second-weight integrals produces new, non-trivial constraints on the
associated theoretical representations.\footnote{As an example, consider the step-two choice $w_2(y)=1-y^2$.
The theory side of the new $w_2$ FESR introduces the single new fit parameter,
$C_6$. If the step-one theory representation is reliable, the difference between
$I_{\rm ex}^{(w_2)}$ and the sum of $D=0$ and DV contributions to
$I_{\rm th}^{(w_2)}$ implied by the step-one $\alpha_s$ and DV-parameter
fit values should scale as $1/s_0^3$. If not, the combined, two-weight fit
will have to shift $\alpha_s$ and the DV parameters away from their step-one
values. Adding the $w_2$ FESR {\it at multiple values of $s_0$} thus provides new
constraints on the theory representations.}

There is, however, a technical complication to be dealt with in order
to access such additional theory-side constraints. Explicitly, for such
two-weight fits, it is impossible to perform a standard $\chi^2$ fit
when, as in most of the fits we consider, we include all $s_0>s_0^{\rm min}$
accessible for our 63-cluster combination. The reason, first pointed out
in Ref.~\cite{PRS}, and explained in further detail in Ref.~\cite{DVvstOPE},
is as follows. Consider two sets of spectral integrals,
$I_{\rm exp}^{(w_n)}(s_0)$ and $I_{\rm exp}^{(w_m)}(s_0)$, sharing a
common set of $N$ such $s_0$, the first involving weight $w_n$ and the
second weight $w_m$. It turns out that, in this case, only $N+1$ of
these $2N$ integrals are statistically independent. The correlation
matrix of the full set is singular, with $N-1$ eigenvalues exactly
equal to zero. This precludes forming the standard $\chi^2$ minimizer.
One is left with two options, either to throw out all but one of the
second set of spectral integrals and perform a standard $\chi^2$ fit
to the resulting reduced $N+1$-member set, or retain
the full $2N$-member set, but use a different, non-standard-$\chi^2$
minimizer. The first option has a major disadvantage, namely that, with the
reduced spectral integral set containing only one $w_m$-weighted spectral
integral, at a single $s_0$, one has no access to the additional
theory constraints an analysis including the $w_m$-weighted spectral
integrals at multiple $s_0$ would have provided. In view of this
limitation, we choose option two, retaining the full sets of both
spectral integrals and working with a different minimizer.

The alternate minimizer we choose to work with is the block-diagonal
quadratic form
\begin{equation}
\label{Q2}
Q^2_{nm}=\c^2(w_n)+\c^2(w_m)\ ,
\end{equation}
with $\c^2(w_k)$, $k=n,m$, the standard $\c^2$ function constructed
from the data for $I_{\rm exp}^{(w_k)}(s_0)$ alone, with its $N\times N$
covariance matrix. Correlations between spectral integrals involving
different weights are not included in constructing $Q^2$, though they
are included when obtaining the errors on and correlations among the
final fit parameter values.

Focusing on the weights considered in this paper, we note that the theory
representation of $I_{\rm exp}^{(w_0)}(s_0)$ is not sensitive to any of the
OPE condensates $C_{D\geq 4}$, while that of $I_{\rm exp}^{(w_2)}(s_0)$ is
sensitive to $C_6$, that of $I_{\rm exp}^{(w_3)}(s_0)$ to $C_6$ and $C_8$,
and that of $I_{\rm exp}^{(w_4)}(s_0)$ to $C_6$ and $C_{10}$. Each of the
corresponding terms in the OPE, Eq.~(\ref{OPE}), has a different $s$ dependence,
and thus produces a different $s_0$ dependence for the associated
contour integral. Adding the two-weight fits based on the minimizers
$Q^2_{02}$, $Q^2_{03}$ and $Q^2_{04}$ thus tests for stability of the
values of $\a_s$, $C_6$ and the DV parameters and, indirectly, the
reliability of the theoretical representations. Note that the theory-side
DV integral contributions have oscillatory $s_0$ dependences. This is
very different from the inverse power dependences, $1/s_0^{D/2}$, of the
integrated dimension-$D$ OPE contributions. With oscillatory behavior
clearly evident in the experimental spectral function data for
$s>1.5$~GeV$^2$ ({\it cf.} Fig.~\ref{finalspec-fig}), both the OPE and
DV parts of our theory representation are tested by these two-weight fits.

As noted above, although the ``block-diagonal'' minimizer defined in
Eq.~(\ref{Q2}) does not incorporate the correlations between spectral integrals,
$I^{(w_n)}(s_0)$ and $I^{(w_m)}(s_0)$, with different weights,
one can still obtain parameter errors for such fits by linear error
propagation including the full two-weight covariance matrix \cite{alphas1},
and associated $p$-values using the method of Ref.~\cite{BS}. In the
fits we report in Sec.~\ref{results} all parameter errors shown for
two-weight fits have been obtained accordingly.

\subsection{\label{results} Results}
In Table~\ref{tabw0} we show, as a function of $s_0^{\rm min}$, the results 
of fits to $I_{\rm exp}^{(w_0)}(s_0)$ for $s_0$ in a range extending from 
$s_0^{\rm min}$ up to the highest $s$ in our 63-cluster set,
$s_{\rm max}=3.0871$~GeV$^2$. We set the parameter $c$ in 
Eq.~(\ref{ansatz}) to zero in this section, but will consider the effect of
non-zero values in Sec.~\ref{analysis} below. The dependence of the fit
results on $s_0^{\rm min}$ for $\a_s(m_\t^2)$ and the DV parameters
are shown by the black points in Figs.~\ref{figalphas} and~\ref{figDV},
respectively.

\begin{figure}[t!]
\begin{center}
\includegraphics*[width=11cm]{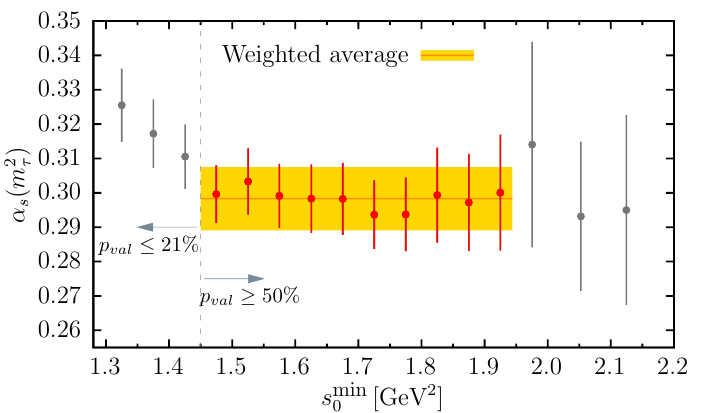}
\end{center}
\vspace*{-4ex}
\begin{quotation}
\floatcaption{figalphas}%
{{\it $\a_s(m_\t^2)$ of Table~\ref{tabw0} as a function of 
$s_0^{\rm min}$ (showing statistical errors only). The yellow area 
correspond to the average reported in Eq.~(\ref{w0pars}); this average is 
computed from the data points indicated in red (see text). The thin vertical 
dashed line separates the regions in which the $p$-values shown in 
Table~\ref{tabw0} are less than 21\% (to the left), from the region where 
they are greater than 50\% (to the right).}}
\end{quotation}
\vspace*{-4ex}
\end{figure}

\begin{figure}[!t]
\begin{center}
\includegraphics*[width=7.3cm]{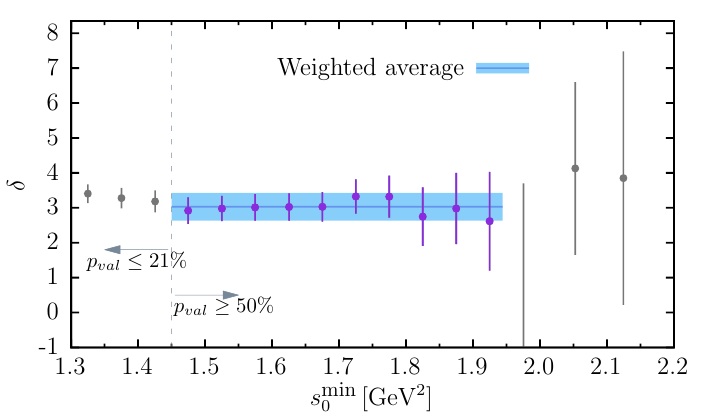}
\hspace{0.2cm}
\includegraphics*[width=7.3cm]{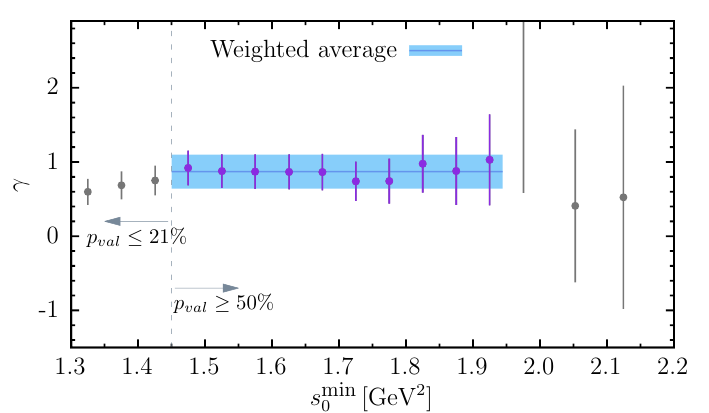}
\vspace{1cm}
\includegraphics*[width=7.3cm]{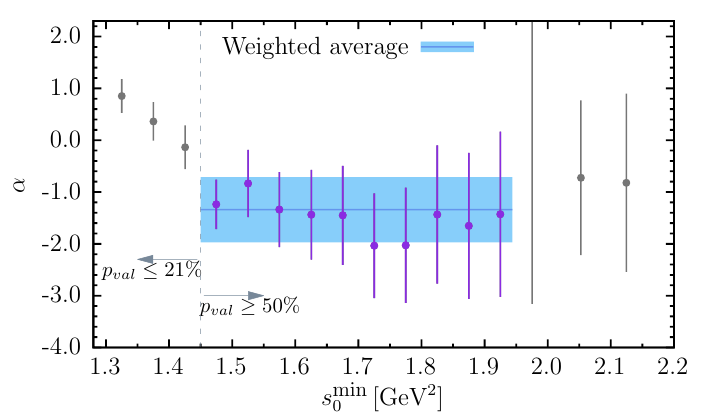}
\hspace{0.2cm}
\includegraphics*[width=7.3cm]{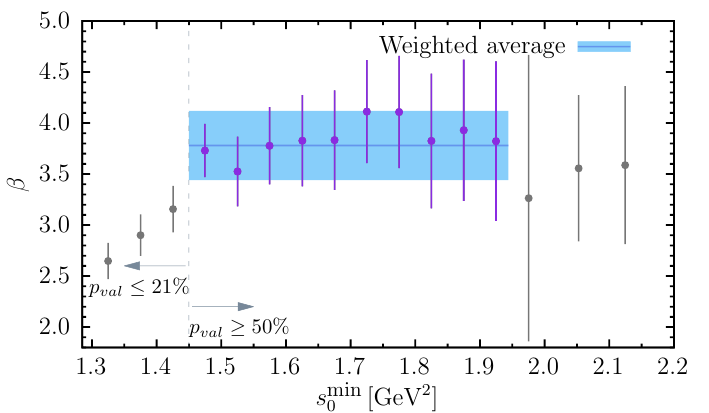}
\end{center}
\vspace*{-8ex}
\begin{quotation}
\floatcaption{figDV}%
{{\it DV parameters of Table~\ref{tabw0} as a function
of $s_0^{\rm min}$ (showing statistical errors only). The blue 
areas correspond to the averages reported in Eq.~(\ref{w0pars}); these 
averages are computed from the data points indicated in purple (see text). 
$\b$ and $\g$ are in {\rm GeV}$^{-2}$. The thin vertical dashed line 
separates the regions in which the $p$-values shown in Table~\ref{tabw0} 
are less than 21\% (to the left), from the region where they are greater 
than 50\% (to the right).}}
\end{quotation}
\vspace*{-4ex}
\end{figure}

Inspecting the table, we see that fits with $s_0^{\rm min}$ ranging from
$1.4747$ to $1.9249$~GeV$^2$ have good $p$-values, with parameter values
stable over the full range. For smaller $s_0^{\rm min}$, the
$p$-values (while acceptable) show a steep drop, and parameter values
(notably $\a_s(m_\t^2)$, $\a$ and $\b$) are less stable. This suggests
that the large-$s$ theoretical representation employed in the
fit starts to break down. For $s_0^{\rm min}$ values greater
than $1.9249$~GeV$^2$, $p$-values remain good, but parameters are less
well determined; parameter errors grow as the number of
available data points for each fit decreases.

\begin{table}[t]
\begin{center}
\begin{tabular}{|l|c|c|l|l|r|r|c|}
\hline
$~~s_0^{\rm min}$ & $\c^2$/dof & $p$-val. &  $~~\a_s(m_\t^2)$ & $~~~~\d$
& $\g~~~~~$ & $\a~~~~~$ & $\b$ \\
\hline
1.3250 & 28.25/18 & 0.058 & 0.3255(107) & 3.41(27) & 0.60(18) & 0.85(32) & 2.65(18) \\
1.3752 & 23.76/17 & 0.13 & 0.3172(100) & 3.28(30) & 0.69(19) & 0.36(37) & 2.90(20) \\
1.4254 & 20.13/16 & 0.21 & 0.3106(95) & 3.18(32) & 0.75(20) & $-$0.14(42) & 3.16(23) \\
1.4747 & 10.39/15 & 0.79 & 0.2996(84) & 2.92(39) & 0.92(24) & $-$1.24(48) & 3.73(26) \\
1.5252 & 9.36/14 & 0.81 & 0.3033(98) & 2.98(37) & 0.88(23) & $-$0.84(65) & 3.53(34) \\
1.5747 & 7.99/13 & 0.84 & 0.2991(95) & 3.01(39) & 0.87(24) & $-$1.34(73) & 3.78(38) \\
1.6254 & 7.95/12 & 0.79 & 0.2983(102) & 3.02(40) & 0.87(24) & $-$1.44(87) & 3.83(45) \\
1.6752 & 7.95/11 & 0.72 & 0.2982(109) & 3.03(43) & 0.86(25) & $-$1.45(97) & 3.83(49) \\
1.7251 & 6.78/10 & 0.75 & 0.2936(103) & 3.33(50) & 0.74(27) & $-$2.0(1.0) & 4.11(52) \\
1.7750 & 6.78/9 & 0.66 & 0.2937(112) & 3.32(61) & 0.74(31) & $-$2.0(1.2) & 4.11(57) \\
1.8250 & 5.61/8 & 0.69 & 0.2994(147) & 2.75(85) & 0.98(39) & $-$1.4(1.4) & 3.82(68) \\
1.8750 & 5.48/7 & 0.60 & 0.2972(152) & 3.0(1.0) & 0.88(46) & $-$1.7(1.5) & 3.93(71) \\
1.9249 & 5.31/6 & 0.50 & 0.3001(181) & 2.6(1.4) & 1.03(62) & $-$1.4(1.7) & 3.82(81) \\
1.9754 & 4.17/5  & 0.53 & 0.3141(186) & 0.7(1.5) & 1.82(64) & $-$0.3(1.8) & 3.26(89) \\
2.0526 & 2.45/4 & 0.65 & 0.2932(247) & 4.1(1.1) & 0.4(1.1) & $-$0.7(1.6) & 3.56(79) \\
2.1246 & 2.44/3 & 0.49 & 0.2950(252) & 3.9(3.0) & 0.5(1.3) & $-$0.8(1.8) & 3.59(84) \\
\hline
\end{tabular}
\end{center}
\floatcaption{tabw0}%
{{\it Results of fits to $I^{(w_0)}_{\rm exp}(s_0)$ employing the combined
63-cluster spectral function, with $s_{\rm max}=3.0871$~{\rm GeV}$^2$,
$s_0^{\rm min}$ in {\rm GeV}$^2$ and $\b$ and $\g$ in {\rm GeV}$^{-2}$.}}
\end{table}

We have checked how the results for $\a_s(m_\t^2)$ change when we use the
spectral function with inflated errors. Over the whole range of
$s_0^{\rm min}$ shown in the table, we find little change in the
errors, and no significant impact on the values, obtained for $\a_s(m_\t^2)$.
More interesting is what happens if we replace the updated hadronic
BFs from Ref.~\cite{hflav22} used in the construction of the spectral function
with those of the earlier 2019 HFLAV compilation, Ref.~\cite{hflav2019}.
With the older BFs, which had smaller errors, especially for the
$\pi^- 3\pi^0$ mode, the errors on $\a_s(m_\t^2)$ for
$s_0^{\rm min}\in[1.4747,1.9249]$~GeV$^2$ become about $6-20\%$
smaller. As a consequence, the error on our new value for $\a_s(m_\t^2)$
turns out to be larger than the error on the value obtained in Ref.~\cite{alphas20}, despite the
addition of the high-precision Belle $2\pi$ data to the analysis.

We have computed the covariance matrices for the values of $\alpha_s$ and 
the DV parameters obtained from fits with different $s_0^{\rm min}$ between
$1.4747$ and $1.9249$~GeV$^2$. For our final $\a_s$ result, we average the
ten values in this range, using a diagonal fit to avoid the possible
bias the strong correlations between them might produce. The error on
this average, however, is determined by linearly propagating the
full $\a_s$ covariance matrix. The same averaging strategy is
used for the DV parameters. We find the values (statistical errors only)
\begin{eqnarray}
\label{w0pars}
\a_s(m_\t^2)&=&0.2983(92)\ ,\\
\d&=&3.03(40)\ ,\nonumber\\
\g&=&0.87(23)\ \mbox{GeV}^{-2}\ ,\nonumber\\
\a&=&-1.34(63)\ ,\nonumber\\
\b&=&3.78(34)\ \mbox{GeV}^{-2}\ .\nonumber
\end{eqnarray}
The final $\a_s(m_\t^2)$ and DV parameter results are shown in yellow in
Fig.~\ref{figalphas} and blue in Fig.~\ref{figDV}, respectively, with 
statistical errors only.
While the errors have been computed using the full covariance matrix
for the ten values of each parameter for $s_0^{\rm min}$ between
$1.4747$ and $1.9249$~GeV$^2$ in Table~\ref{tabw0}, we note
that taking the errors for the fit with the largest $p$-value in the table leads to reasonable
estimates of the average parameter errors.
We will take Eq.~(\ref{w0pars}) to define our central results and statistical errors. The
systematic error on $\a_s(m_\t^2)$ will be discussed in the following 
subsection.

\begin{figure}[t]
\begin{center}
\includegraphics*[width=7.4cm]{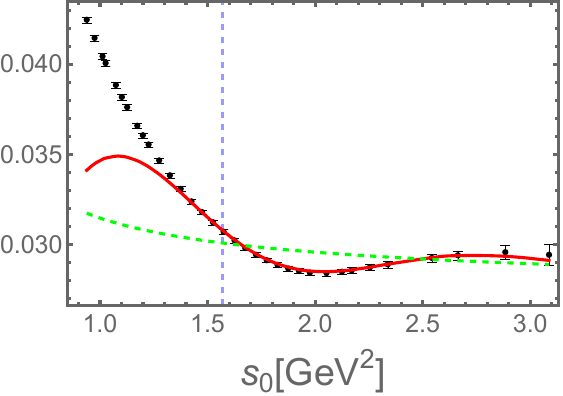}
\hspace{0.3cm}
\includegraphics*[width=7cm]{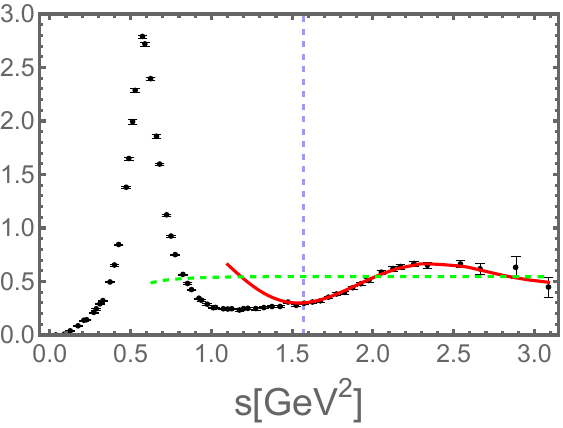}
\end{center}
\begin{quotation}
\floatcaption{fits}%
{{\it Left panel: Comparison of the fitted theoretical representations
and the spectral moments $I^{(w_0)}_{\rm exp}(s_0)$ from the fit with
$s_0^{\rm min}=1.5747$~{\rm GeV}$^2$.
Right panel: Comparison of $2\pi^2$ times the 63-cluster experimental
spectral function and theoretical representation thereof resulting
from the fit. The black symbols denote the experimental results,
the solid red curves the fitted theory results, and the green dashed
curves the OPE parts of the fitted theory curves. The vertical blue
dashed lines mark $s=s_0^{\rm min}=1.5747$~{\rm GeV}$^2$.}}
\end{quotation}
\vspace*{-4ex}
\end{figure}

For illustrative purposes, the fit to $I^{(w_0)}_{\rm exp}(s_0)$ for 
$s_0^{\rm min}=1.5747$~GeV$^2$ is
displayed in the left panel of Fig.~\ref{fits}. The right panel of the same
figure shows a comparison of the representation for $\rho_{ud;V}(s)$ obtained
using the parameters of this fit with the combined experimental result
obtained in Sec.~\ref{datasec}.

In Table~\ref{tabw0w2}, we present the results for the block-diagonal
simultaneous fits to $I^{(w_0)}_{\rm exp}(s_0)$ and
$I^{(w_2)}_{\rm exp}(s_0)$, limiting ourselves to the ten
values of $s_0^{\rm min}$ used in obtaining the ten-point
averages in Eq.~(\ref{w0pars}) above. Taking again diagonal averages
of the ten values from this interval, we find the results
\begin{eqnarray}
\label{w2pars}
\a_s(m_\t^2)&=&0.2961(91)\ ,\\
\d&=&2.73(40)\ ,\nonumber\\
\g&=&1.05(25)\ \mbox{GeV}^{-2}\ ,\nonumber\\
\a&=&-1.65(65)\ ,\nonumber\\
\b&=&3.95(35)\ \mbox{GeV}^{-2}\ ,\nonumber\\
C_6&=&-0.0084(16) \ \mbox{GeV}^6\ .\nonumber
\end{eqnarray}
We have chosen the errors for the fit with the largest $p$-value
as the errors for each parameter in
Eqs.~(\ref{w2pars}). As we will not use the two-weight fits
for our central results, this choice is adequate for
checking the stability of our fits. The first five parameter values,
especially $\a_s(m_\t^2)$, are in good agreement with those in
Eq.~(\ref{w0pars}), while $C_6$ appears here for the first time.

\begin{table}[t!]
\small
\begin{center}
\vspace*{4ex}
\begin{tabular}{|c|c|c|c|c|c|c|c|c|}
\hline
$~~s_0^{\rm min}$ & $Q_{02}^2$/dof & $p$-val. &  $~~\a_s(m_\t^2)$ & $~~~~\,\d$ & $\g$ & $\a$ & $\b$ & $10^2 C_6$ \\
\hline
1.4747 & 33.96/34 & $0.31$ & 0.2976(83) & 2.64(40) & 1.09(24) & $-$1.43(48) & 3.83(27)& $-$0.79(13) \\
1.5252 & 30.97/32 & $0.33$ & 0.2992(91) & 2.72(40) & 1.04(25) & $-$1.32(65) & 3.78(35)& $-$0.77(16) \\
1.5747 & 29.11/30 & $0.32$ & 0.2960(90) & 2.73(41) & 1.04(25) & $-$1.76(73) & 4.00(39)& $-$0.85(15)  \\
1.6254 & 28.96/28 & $0.25$  & 0.2954(95) & 2.76(42) & 1.03(25) & $-$1.86(89) & 4.05(46) & $-$0.86(17) \\
1.6752 & 28.23/26 & $0.21$  & 0.2947(98) & 2.83(44) & 1.00(26) & $-$1.99(97) & 4.11(50)& $-$0.89(17) \\
1.7251 & 27.03/24 & $0.18$  & 0.2919(97) & 3.01(52) & 0.92(28) & $-$2.39(1.06) & 4.30(54) & $-$0.95(17) \\
1.7750 & 26.60/22 & $0.13$  & 0.2932(109) & 2.84(66) & 1.00(33) & $-$2.21(1.24) & 4.21(62)& $-$0.92(21)  \\
1.8250 & 24.41/20 & $0.12$ & 0.2979(140) & 2.30(89) & 1.22(41) & $-$1.64(1.53) & 3.94(76)& $-$0.79(33) \\
1.8750 & 24.40/18 & $0.08$ & 0.2978(153) & 2.3(1.2) & 1.22(55) & $-$1.7(1.7) & 3.95(85)& $-$0.80(38)\\
1.9249 & 22.61/16 & $0.07$ & 0.3034(195) & 1.3(1.6) & 1.66(68) & $-$1.2(2.2) & 3.7(1.1)& $-$0.60(57) \\
\hline
\end{tabular}
\end{center}
\floatcaption{tabw0w2}%
{{\it Results of combined fits to $I^{(w_0)}_{\rm exp}(s_0)$ and
$I^{(w_2)}_{\rm exp}(s_0)$ employing our 63-cluster spectral
function, with $s_{\rm max}=3.0869$~{\rm GeV}$^2$, $s_0^{\rm min}$ in
{\rm GeV}$^2$, $\b$ and $\g$ in {\rm GeV}$^{-2}$ and $C_6$ in {\rm GeV}$^6$.}}
\end{table}

\begin{table}[t!]
{\footnotesize
\begin{center}
\vspace*{4ex}
\begin{tabular}{|c|c|c|c|c|c|c|c|c|c|}
\hline
$~~s_0^{\rm min}$ & $Q_{03}^2$/dof & $p$-val. &  $~~\a_s(m_\t^2)$ & $~~~~\,\d$ & $\g$ & $\a$ & $\b$ & $10^2 C_6$ & $10^2 C_8$ \\
\hline
1.4747 & 3.22/7 & $0.35$ & 0.3053(96) & 3.22(40) & 0.72(25) & $-$1.11(60) & 3.66(32) & $-$0.71(16) & 1.15(27) \\
1.5252 & 8.14/7 & $0.07$ & 0.2999(90) & 3.06(41) & 0.83(25) & $-$1.75(71) & 4.00(38) & $-$0.84(14) & 1.40(25) \\
1.5747 & 1.95/5 & $0.24$ & 0.2997(92) & 3.27(39) & 0.71(23) & $-$1.79(75) & 4.01(39)& $-$0.84(14) & 1.40(25)  \\
1.6254 & 1.65/5 & $0.27$  & 0.3001(103)  & 3.40(44) & 0.65(26) & $-$1.71(88) & 3.96(45)  & $-$0.83(17) & 1.41(31)  \\
1.6752 & 3.63/5 & $0.10$  & 0.2901(88) & 3.41(48) & 0.69(27) & $-$3.18(95)  & 4.70(48)  & $-$1.04(12) & 1.81(22)\\
1.7251 & 0.75/3 & $0.001$  & 0.2927(95) & 3.75(64) & 0.50(33) & $-$2.7(1.0) & 4.45(50) & $-$0.99(15) & 1.71(28)  \\
1.7750 & 1.04/3 & $0.003$  & 0.3085(181) & 2.9(1.1) & 0.85(49) & $-$0.9(1.5) & 3.54(75) & $-$0.63(44)  & 0.99(91) \\
1.8250 & 1.01/3 & $0.02$ & 0.3104(257) & 1.5(1.7) & 1.53(74) & $-$0.7(2.5) & 3.5(1.2) & $-$0.52(76) & 0.6(1.8) \\
\hline
\end{tabular}
\end{center}
\floatcaption{tabw0w3}%
{{\it Results of combined fits to $I^{(w_0)}_{\rm exp}(s_0)$
and $I^{(w_3)}_{\rm exp}(s_0)$ employing our 63-cluster spectral
function,  with $s_{\rm max}=3.0869$~{\rm GeV}$^2$, $s_0^{\rm min}$ in
{\rm GeV}$^2$, $\b$ and $\g$ in {\rm GeV}$^{-2}$, $C_6$ in {\rm GeV}$^6$
and $C_8$ in {\rm GeV}$^8$.
In the fits, the data for $I^{(w_0)}_{\rm exp}(s_0)$ and
$I^{(w_3)}_{\rm exp}(s_0)$ have been thinned by a factor 3.}}}
\end{table}

\begin{table}[t!]
{\footnotesize
\begin{center}
\vspace*{4ex}
\begin{tabular}{|c|c|c|c|c|c|c|c|c|c|}
\hline
$s_0^{\rm min}$ & $Q^2_{04}$/dof & $p$-val. &  $\a_s(m_\t^2)$ & $\d$ & $\g$
& $\a$ & $\b$ & $10^2 C_6$ & $10^2 C_{10}$\\
\hline
1.4747 & 2.96/7 & $0.39$ & 0.3052(96) & 3.22(40) & 0.72(25) & $-$1.10(60) & 3.66(32) & $-$0.69(16) & $-$1.33(42) \\
1.5252 & 8.02/7 & $0.07$ & 0.2998(91) & 3.07(40) & 0.82(25) & $-$1.72(71) & 3.99(38) & $-$0.82(15) & $-$1.75(42) \\
1.5747 & 1.72/5 & $0.27$ & 0.2996(93) & 3.27(39) & 0.71(23) & $-$1.77(76) & 4.00(40)& $-$0.82(15) & $-$1.79(43)  \\
1.6254 & 1.50/5 & $0.29$  & 0.3000(103)  & 3.40(44) & 0.65(26) & $-$1.69(88) & 3.95(45)  & $-$0.82(18) & $-$1.81(53)  \\
1.6752 & 3.58/5 & $0.11$  & 0.2900(88) & 3.41(48) & 0.69(27) & $-$3.16(95)  & 4.69(48)  & $-$1.03(13) & $-$2.54(40)\\
1.7251 & 0.61/3 & $0.002$  & 0.2926(96) & 3.72(65) & 0.51(33) & $-$2.7(1.0) & 4.44(50) & $-$0.98(15) & $-$2.38(52)  \\
1.7750 & 0.85/3 & $0.008$  & 0.3091(189) & 2.9(1.1) & 0.88(49) & $-$0.8(1.6) & 3.50(77) & $-$0.59(48)  & $-$0.9(1.9) \\
1.8250 & 0.91/3 & $0.03$ & 0.3112(266) & 1.4(1.8) & 1.56(75) & $-$0.6(2.5) & 3.4(1.3) & $-$0.47(82) & 0.2(4.4) \\
\hline
\end{tabular}
\end{center}
\floatcaption{tabw0w4}%
{{\it Results of combined fits to $I^{(w_0)}_{\rm exp}(s_0)$ and
$I^{(w_4)}_{\rm exp}(s_0)$ employing the 63-cluster spectral function,
with $s_{\rm max}=3.0869$~{\rm GeV}$^2$, $s_0^{\rm min}$ in {\rm GeV}$^2$,
$\b$ and $\g$ in {\rm GeV}$^{-2}$, $C_6$ in {\rm GeV}$^6$ and $C_{10}$ in
{\rm GeV}$^{10}$.
In the fits, the data for $I^{(w_0)}_{\rm exp}(s_0)$ and
$I^{(w_4)}_{\rm exp}(s_0)$ have been thinned by a factor 3.}}}
\end{table}

As in Ref.~\cite{alphas20}, in the case of simultaneous block-diagonal fits
to $I^{(w_0)}_{\rm exp}(s_0)$ and either $I^{(w_3)}_{\rm exp}(s_0)$ or
$I^{(w_4)}_{\rm exp}(s_0)$, we find that the correlation  matrices for
the spectral moments with the doubly pinched weights $w_n$, with
$n=3$ or $4$, have very small eigenvalues, around $10^{-10}$, three
orders of magnitude smaller than the smallest eigenvalue for the set of
$I^{(w_2)}_{\rm exp}(s_0)$ or $I^{(w_0)}_{\rm exp}(s_0)$ integrals.
We find that if we ``thin'' the set of integrals used in the fit,
starting at a given $s_0^{\rm min}$ and including only every second,
third, \etc, of the available higher $s_0$, the $Q^2/$dof
values drop rapidly to a value below $1$, and the fit stabilizes
as we increase the degree of thinning. Tables~\ref{tabw0w3} and \ref{tabw0w4}
show the results of these fits for the cases $n=3$ and $n=4$, thinning the
data in these cases by a factor three.

In the tables for the two-weight fits involving the doubly pinched weights, 
we see that the results fluctuate a little more than in the corresponding 
parts of Tables~\ref{tabw0} and \ref{tabw0w2}. The reason is that, because 
of the thinning, the sets of $s_0$ values used for the three fits starting 
at, for example, $s_0^{\rm min}=1.5747$, $1.6254$ and $1.6752$~GeV$^2$
are different. The integrals $I^{(w_{3,4})}_{\rm exp}(s_0=1.6254)$ and
$I^{(w_{3,4})}_{\rm exp}(s_0=1.6752)$ are not part of the fit starting with
$I^{(w_{3,4})}_{\rm exp}(s_0=1.5747)$ whereas, if the data were
not thinned, they would be.\footnote{Of course, this does {\it not}
mean that the data used in these three fits are fully independent,
because all $I^{(w)}_{\rm exp}(s_0)$ are strongly correlated, especially
for nearby $s_0$ values.}

Because of the thinning, the fits for $s_0^{\rm min}=1.7251$, $1.7750$ and
$1.8250$~GeV$^2$ are based on very few degrees of freedom, and thus tend 
to be less stable; $p$-values for these fits drop steeply from those for fits with 
smaller $s_0^{\rm min}$ values. Fits with $s_0^{\rm min}=1.8750$ GeV$^2$
and higher have only one degree of freedom, and are very unstable. We 
thus take the parameter averages only over the fits for the first five 
$s_0^{\rm min}$ values shown in the tables, and obtain
\begin{equation}
\label{w0w2w3w4pars}
\begin{array}{crrrr}
 & w_0 & w_0\ \&\ w_2 &  w_0\ \&\ w_3 & w_0\ \&\ w_4  \\
& & & & \\
\a_s(m_\t^2) & 0.2983(92) & 0.2961(91) & 0.2987(96) & 0.2986(96) \\
\d & 3.03(40) & 2.73(40) & 3.26(40) & 3.26(40) \\
\g & 0.87(23) & 1.05(25) & 0.72(25) & 0.72(25) \\
\a & -1.34(63) & -1.65(65) & -1.73(60) & -1.71(60) \\
\b & 3.78(34) & 3.95(35) & 3.99(32) & 3.98(32) \\
C_6 & & -0.0084(16) & -0.0087(16) & -0.0085(16) \\
C_8 & & & 0.0147(27) & \\
C_{10} & & & & -0.0186(42) \\
\end{array}
\end{equation}
In this equation, we have, for ease of comparison also listed the
values obtained in Eqs.~(\ref{w0pars}) and~(\ref{w2pars}). We recall that
the errors in the $w_0$ column have been obtained from the full covariance
matrix for each parameter, while those in the other three columns are
rough estimates obtained by taking the errors for the fit with the
largest $p$-value in each of the tables.

Comparing the parameter values in Eq.~(\ref{w0w2w3w4pars}), we see that
the values obtained are consistent across all these fits, in
particular for the OPE parameters $\a_s(m_\t^2)$ and $C_6$. The DV
parameter values for the two-weight fits agree with the
central values  of the $w_0$ fit, within the errors on that fit.

\subsection{\label{analysis} Analysis and comparison with Ref.~\cite{alphas20}}
We begin with a comparison between our new result
\begin{equation}
\label{w0valrepeat}
\a_s(m_\t^2)=0.2983(92)_{\rm stat} \ ,
\end{equation}
given in Eq.~(\ref{w0pars}) and that obtained in Ref.~\cite{alphas20}, 
which is (showing the statistical error only)
\begin{equation}
\label{as20}
\a_s(m_\t^2)=0.3077(65)_{\rm stat}\ .
\end{equation}
The new central value is lower, and the new error larger. In order to 
probe possible reasons for these differences, we carried out several 
other fits. First, we considered what happens if the Belle 
$2\p$ data are omitted from the combination, and the BFs of Ref.~\cite{hflav2019} 
are used instead of the updated ones of Ref.~\cite{hflav22}. This allows for a 
direct comparison with the result from Ref.~\cite{alphas20}, which also
employed only ALEPH and OPAL data and HFLAV 2019 BF input, but used a
different data combination algorithm. This yields
\begin{equation}
\label{asnoBelleBF19}
\a_s(m_\t^2)=0.3075(84)_{\rm stat} \qquad \mbox{2019 BFs; without 
Belle data}\ .
\end{equation}
The central value is very close to that of Eq.~(\ref{as20}), but the error is
larger.\footnote{Also the DV parameters from this fit are in excellent 
agreement with those of Ref.~\cite{alphas20}.} This implies that the new 
data-combination algorithm leads to somewhat larger statistical errors 
than that of Eq.~(\ref{as20}). Of course, with only the $\p^-\p^0$ mode 
available from Belle, we are forced to use the new algorithm if 
the goal is to include the Belle data. We also note that our 
treatment of the $4\p$ modes is less reliant on the poorly determined
off-diagonal covariances of the ALEPH and OPAL $4\pi$ data than was that 
of Ref.~\cite{alphas20}, with the OPAL $\p^-3\p^0$ data now included through a 
diagonal fit.  This also leads to some increase in the statistical error on $\a_s(m_\t^2)$.

Next, we considered what happens if we add the Belle $2\p$ data, 
keeping BFs from Ref.~\cite{hflav2019} instead of Ref.~\cite{hflav22}. This yields
\begin{equation}
\label{asoldBF}
\a_s(m_\t^2)=0.3055(81)_{\rm stat} \qquad \mbox{2019 BFs; with 
Belle data}\ .
\end{equation}
Finally, we also give the result obtained without Belle 
data, but using the 2022 BFs of Ref.~\cite{hflav22}:
\begin{equation}
\label{asnoBelle}
\a_s(m_\t^2)=0.3001(97)_{\rm stat} \qquad \mbox{2022 BFs; without 
Belle data}\ .
\end{equation}
Comparing Eqs.~(\ref{w0pars}),~(\ref{asnoBelleBF19}),~(\ref{asoldBF}) 
and~(\ref{asnoBelle}) we see that the use of the new BFs has two 
effects: central values with the new HFLAV BFs are clearly smaller, and 
errors are larger. The latter is no surprise since the updated BF 
$B_{\p^-3\p^0}$ has an error a factor of $2.4$ larger. We also see 
that including the Belle data somewhat lowers the central value 
for $\a_s(m_\t^2)$, but this effect is much smaller than the impact of the 
switch from 2019 to 2022 HFLAV BFs. We conclude that the change from the 
2020 result in Eq.~(\ref{as20}) to the new result in Eq.~(\ref{w0valrepeat}) is caused 
by changes in the input data, and not by a change in our methodology, 
witness the good agreement between Eq.~(\ref{as20}) and Eq.~(\ref{asnoBelleBF19}).
A visual summary of the above explorations is provided in
Fig.~\ref{alphasMZvariousinputs}, which displays the results
obtained for $\alpha_s$ using the various experimental and BF inputs, 
and data combination methodologies, run up to the $n_f=5$ scale $\mu =m_Z$.

\begin{figure}[t]
\begin{center}
\includegraphics*[width=11cm]{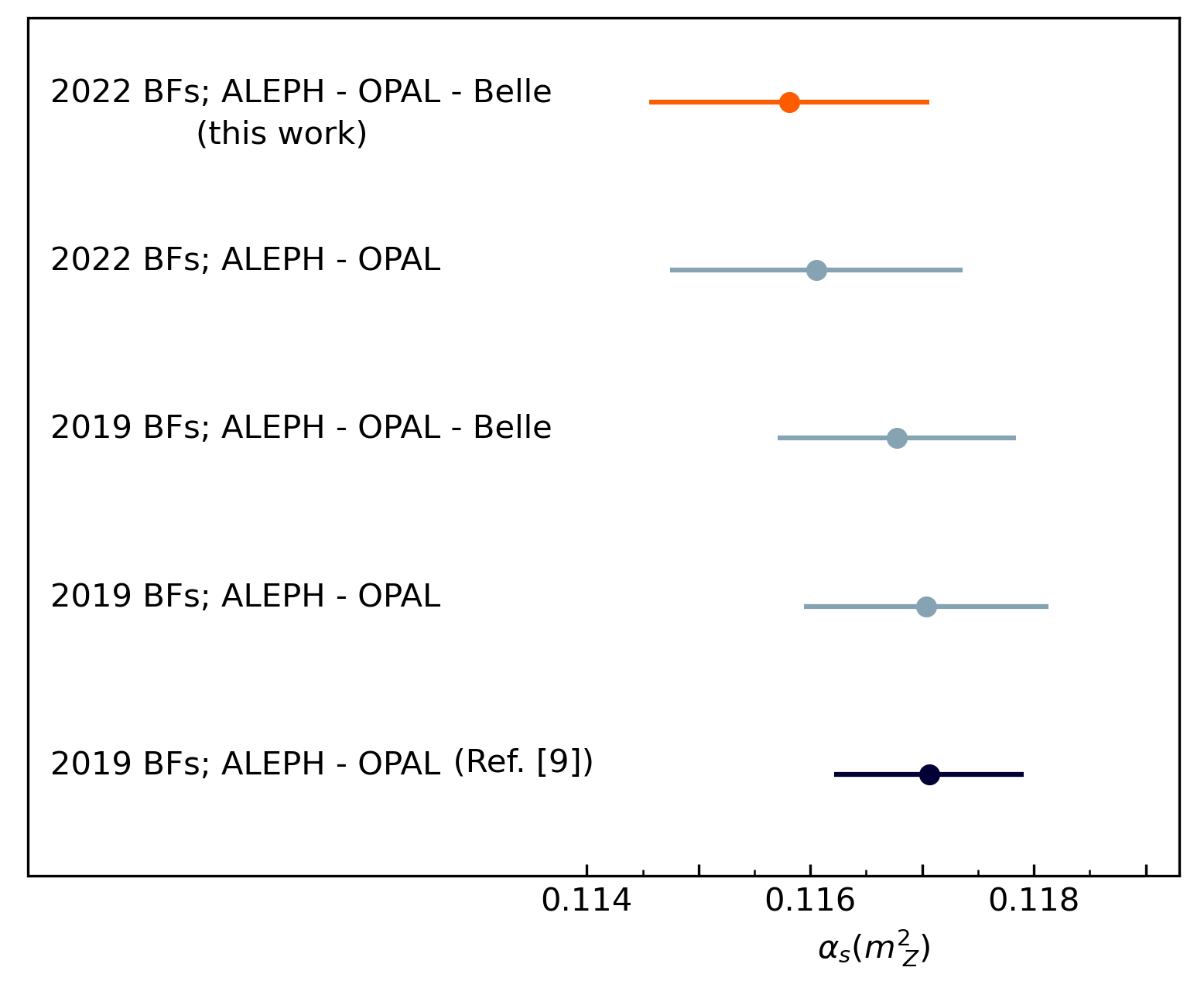}
\end{center}
\begin{quotation}
\floatcaption{alphasMZvariousinputs}%
{{\it Results for $\alpha_s(m_Z^2)$ using different choices for the
$2\pi$ data sets included, the input $2\pi$ and $4\pi$ BFs, and
the data combination methodology employed.
All results except the black point are from this work;
the black point is the final result from Ref.~\cite{alphas20}.
Errors shown are statistical only.}}
\end{quotation}
\vspace*{-4ex}
\end{figure}

Next, we turn to a discussion of systematics. The error on $\a_s(m_\t^2)$ 
in our central determination, Eq.~(\ref{w0pars}) is statistical only. This
leaves several systematic effects to be accounted for. First, we take 
the spread of values in Eq.~(\ref{w0w2w3w4pars}), $0.0026$, as our measure of 
the fit-dependence systematic error. We also, as in Ref.~\cite{alphas20}, 
vary the perturbative coefficient $c_{51}$ by $\pm 50\%$ ({\it cf.} 
Sec.~\ref{theory}). This leads to a (symmetrized) variation of $\a_s(m_\t^2)$ 
of $\pm 0.0022$ (at $s_0^{\rm min}=1.5747$~GeV$^2$), which we treat as 
an additional theory error.\footnote{As in Ref.~\cite{alphas20}, 
alternate error estimates based on removing order-$\a_s^5$ terms (\ie, 
setting $c_{5m}=0$), or removing both order-$\a_s^4$ and order-$\a_s^5$ 
terms (\ie, setting both $c_{4m}=0$ and $c_{5m}=0$) lead to differences 
equal to or smaller than the differences obtained from the $50\%$ variation 
in $c_{51}$.} We also considered the variation obtained from choosing 
the scale $\m^2$ in Eq.~(\ref{pertth}) equal to either $2s_0$ or $m_\t^2$, 
instead of $s_0$. The larger of the two resulting scale-change 
variations, $0.0021$, which comes from the choice $\m^2=2s_0$, is (slightly) 
smaller than the estimate obtained from varying $c_{51}$. We thus take 
the larger, $\pm 0.0022$, value as our estimate for the systematic error 
resulting from the truncation of perturbation theory.

As discussed in Ref.~\cite{alphas20}, since the different weights have 
different levels of pinching, and thus different levels of DV suppression, 
as well as different weightings of neglected $\alpha_s$-suppressed 
logarithmic corrections to higher-dimension OPE contributions, the 
variation of the central values for $\a_s(m_\t^2)$ shown in 
Eq.~(\ref{w0w2w3w4pars}) can be taken as an indication of the uncertainties 
resulting from possible shortcomings in the modeling of NP effects.
This variation, $0.0026$, is only $0.9\%$ of the value of 
$\a_s(m_\t^2)$ in Eq.~(\ref{w0pars}) and much smaller than the total
statistical error of $3.1\%$. 

Possible corrections to the DV parametrization in Eq.~(\ref{ansatz}), as well
as logarithmic corrections to the OPE coefficients, were analyzed in
Ref.~\cite{DVvstOPE} and found to be safely negligible for the $\a_s$
determination based on the spectral function of Ref.~\cite{alphas20}. That 
study, following Ref.~\cite{BCGMP}, investigated the correction of order $1/s$ in the DV 
{\it ansatz}~(\ref{ansatz}), 
finding that, while the single-weight, $w_0$ FESR yielded $\chi^2/$dof 
and $\alpha_s$ essentially constant over a wide range of $c>0$, the 
$\chi^2/$dof of the fit and stability of the $\alpha_s$ determination
both deteriorated rapidly for $c$ more negative than about $-1$ GeV$^2$. 
A rapid deterioration was also found in the consistency of the $w_0$ and 
$w_2$ FESRs as $c$ became more negative than $-1$ 
GeV$^2$~\cite{DVvstOPE}.{\footnote{For further details on these
points, see Appendix C of Ref.~\cite{DVvstOPE}.}} The results of 
fits to the spectral integrals, $I^{(w_0)}_{\rm exp}(s_0)$, constructed 
from the new spectral function, with $s_0^{\rm min}=1.5747$~GeV$^2$ and 
variable $c$, are shown in Fig.~\ref{ccorr}. The left panel shows the 
$\c^2$ values and the right panel the $\a_s(m_\t^2)$ values, both as a 
function of $c$.{\footnote{For analogous results produced using the 
spectral function of Ref.~\cite{alphas20}, see Fig.~2 of Ref.~\cite{DVvstOPE}.}}
Taking $c$ negative is seen to lead to a rapid deterioration of the fit 
quality, while $\a_s(m_\t^2)$ is essentially independent of $c$ for positive 
$c$, even for values well outside the range for which the term $c/s$ in 
Eq.~(\ref{ansatz}) is a small correction. The conclusions regarding
such a possible sub-leading correction to our DV {\it ansatz} are thus
the same as those found previously in Ref.~\cite{DVvstOPE} using the spectral 
function of Ref.~\cite{alphas20}. Although the results of fits using combinations of weights with different
degrees of pinching, and hence different weightings of DV contributions,
show, at the current level of precision, no evidence for a significant systematic
uncertainty resulting from choosing $c=0$, we, nonetheless, assign a
systematic error to this choice, obtained by varying $c$ over the range
$\pm 0.8$ GeV$^2$, which corresponds to a $c/s$ correction in Eq.~(\ref{ansatz}) of about $50\%$ at 
the lowest value of $s_0^{\rm min}$ used in the determination of 
$\a_s(m_\t^2)$. This leads to an additional systematic error of $-0.0010$ 
for $c>0$ and $+0.0039$ for $c<0$.

\begin{figure}[t]
\begin{center}
\includegraphics*[width=7cm]{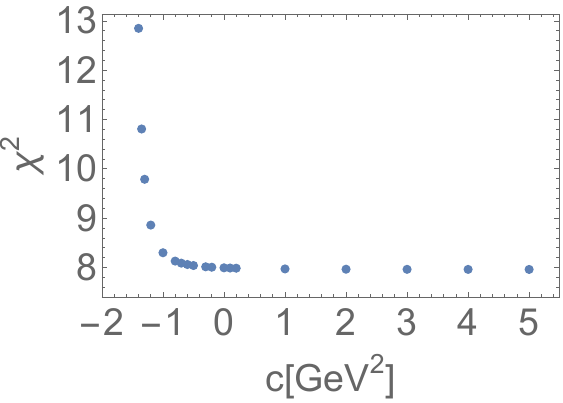}
\hspace{0.3cm}
\includegraphics*[width=7cm]{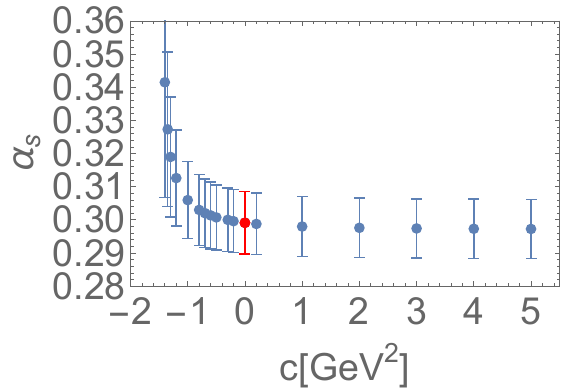}
\end{center}
\begin{quotation}
\floatcaption{ccorr}%
{{\it The results of fits at $s_0^{\rm min}=1.5747~{\rm GeV}^2$ using
the ansatz~(\ref{ansatz}) for DVs, as a function of the parameter $c$
(in ${\rm GeV}^2$). The left and right panels show the dependences
of $\c^2$ and $\a_s(m_\t^2)$ on $c$, respectively. The red point in the
right panel corresponds to the value of $\a_s(m_\t^2)$ in Table~\ref{tabw0}.}}
\end{quotation}
\vspace*{-4ex}
\end{figure}

Combining all errors leads to our final result:
\begin{eqnarray}
\label{final}
\a_s(m_\t^2)&=&0.2983\pm 0.0092_{\rm stat}\pm 0.0026_{\rm fit}
\pm 0.0022_{\rm pert}\pm 0.0025_c\\
&=&0.2983\pm 0.0101\qquad (n_f=3)\ ,\nonumber
\end{eqnarray}
where we symmetrized the error due to the $c/s$ correction in Eq.~(\ref{ansatz}).\footnote{Without
the $c/s$-correction error, the total error would be $\pm0.0098$.}
This can be compared with our final result from Ref.~\cite{alphas20},
\begin{eqnarray}
\label{final20}
\a_s(m_\t^2)&=&0.3077\pm 0.0065_{\rm stat}\pm 0.0038_{\rm pert}\\
&=&0.3077\pm 0.0075\ .\nonumber
\end{eqnarray}
Note that the earlier, 2020 value does not include an estimate
of the uncertainty associated with possible sub-leading corrections
to the leading DV {\it ansatz} since no study involving variation
of the sub-leading contribution parameter, $c$, was considered in 
Ref.~\cite{alphas20}.\footnote{Including the $\pm 0.0025$ version
of the estimate for this uncertainty obtained above using the new
spectral function would increase the error on this earlier determination
from $\pm 0.0075$ to about $\pm 0.0080$.} The errors in Eqs.~(\ref{final}) 
and~(\ref{final20}) are, of course, not uncorrelated, so the two results 
should not be interpreted as agreeing within errors. The reason for the
difference, as explained above, is the change in input data, notably 
the update of the $2\p$ and $4\p$ BFs by HFLAV.

Running the result of Eq.~(\ref{final}) to the $Z$-mass scale using the
standard self-consistent combination of five-loop running \cite{5loop,5loop2}
with four-loop matching \cite{ScSt,CKS} at the charm and bottom thresholds
($2m_c(m_c)$ and $2m_b(m_b)$, respectively, with $\overline{\rm MS}$ masses
from the PDG~\cite{PDG}), we obtain the corresponding $n_f=5$ result
\begin{equation}
\label{asZ}
\a_s(m_Z^2)=0.1159\pm 0.0014\qquad(n_f=5)\ .
\end{equation}
With five-loop running and four-loop matching the uncertainty due to the
running is very small.\footnote{An an example, performing the
matching at $m_c(m_c)$ and $m_b(m_b)$ rather than $2m_c(m_c)$ and
$2m_b(m_b)$ produces a shift of just 0.00008, which does not contribute
to the final uncertainty.}

\section{\label{conclusion} Conclusion}

Our goal in this paper was to include more of the world's data for hadronic
$\t$ decays into the $\t$-based determination of the strong coupling $\a_s$
from the inclusive, non-strange vector-isovector hadronic spectral function.
Previously, we combined ALEPH \cite{ALEPH13} and OPAL \cite{OPAL} data for 
the decays $\t\to\p^-\p^0\n_\t$, $\t\to 2\p^-\p^+\p^0\n_\t$ and 
$\t\to\p^-3\p^0\n_\t$, supplemented with mostly electroproduction 
data and CVC\footnote{Together with $K\bar{K}$ $\t$-decay mode 
distribution data measured by BaBar \cite{babarkkbartau18}.} to 
obtain the residual-mode contributions \cite{alphas20}. As we first 
combined the $2\p$ and $4\p$ data for each of these two experiments before 
combining the resulting single-experiment sums, this did not allow for 
the inclusion of $\p^-\p^0$ data from Belle \cite{Belle} and 
CLEO \cite{CLEO,CLEO2pi}, since no $4\p$-mode results are available
from those experiments. In this paper we therefore changed our algorithm 
for combining data from different experiments to first perform 
single-mode, all-experiment combinations of the $2\p$ data and $4\p$ 
data separately, thus allowing us to include Belle and CLEO data in the 
$2\p$ combination as well as in the construction of the inclusive, 
non-strange vector-isovector hadronic spectral function, $\r_{ud;V}(s)$. 
We found that inclusion of the Belle data has a significant 
impact on the combined spectral function. In contrast, as discussed 
in App.~\ref{CLEOsec}, the impact of including the CLEO data is small and, 
since no complete information is available on the CLEO systematic errors, 
we did not include the CLEO data in the construction of our final result 
for $\r_{ud;V}(s)$.

In addition to including hadronic $\t$-decay data from more experiments, 
we have also updated the values for the input $\p^-\p^0$, $2\p^-\p^+\p^0$ 
and $\p^-3\p^0$ branching fractions with those of Ref.~\cite{hflav22}. This 
has a significant effect on the precision of $\r_{ud;V}(s)$ and
the strong coupling since the new HFLAV value for the $\p^-3\p^0$ 
branching fraction is significantly lower, and has a much larger error 
than, the earlier values reported in Ref.~\cite{hflav2019}. In contrast, 
there have been no relevant improvements to the electroproduction data 
used for the residual-mode input in Ref.~\cite{alphas20}, and we thus carried 
over the residual-mode distributions unchanged from Ref.~\cite{alphas20}.

The main results of our paper are a new determination of the inclusive, 
non-strange vector-isovector hadronic spectral function from the world's 
available data for hadronic $\t$ decays, and a new determination of 
$\a_s(m_\t^2)$ from this spectral function through finite-energy sum rules.
The new spectral function is tabulated in Table~\ref{tabspec}, and shown 
in Fig.~\ref{finalspec-fig}. Our new result for the $n_f=3$ coupling, 
$\a_s(m_\t^2)$, can be found in Eq.~(\ref{final}), and the conversion 
of this result to the corresponding $n_f=5$ value at the $Z$ mass 
in Eq.~(\ref{asZ}). Our results are consistent with those of Ref.~\cite{alphas20}, 
though it should be kept in mind that the current and Ref.~\cite{alphas20} 
results are correlated, as they are based on overlapping data sets. 
The new central value of $\a_s(m_\t^2)$ is $0.0094$ lower than that 
of Ref.~\cite{alphas20}. As can be seen by comparing Eq.~(\ref{w0valrepeat}) with 
Eqs.~(\ref{asoldBF}) and~(\ref{asnoBelle}), almost $80\%$ of this downward shift is
due to the HFLAV update of the $2\p$ and $4\p$ BFs, with roughly $20\%$ 
due to the inclusion of the Belle $2\p$ data. We note, in particular, that 
applying our new data combination method to the data used in Ref.~\cite{alphas20} 
leads to a result fully compatible with that of Ref.~\cite{alphas20}, as a 
comparison of Eqs.~(\ref{as20}) and~(\ref{asnoBelleBF19}) shows.

We also note that the statistical error (and thus, since the 
systematic error is basically unchanged, the combined error) on our new 
result for $\a_s(m_\t^2)$ is larger than in Ref.~\cite{alphas20}. The main reason 
for this increase is again the inclusion of new experimental information:
the Belle $2\p$ data, and the updated branching fractions. While 
the Belle data in principle add significantly more information 
about the $2\p$ mode, inclusion of these data necessitated a change in the 
combination algorithm which turned out to produce an increase in the
error on $\a_s$. The changes in the new algorithm also include a 
treatment of the $4\p$ data which is less reliant on poorly determined ALEPH 
and OPAL $4\pi$ correlations. As can be seen by comparing Eq.~(\ref{as20}) and 
Eq.~(\ref{asnoBelleBF19}), the change in algorithm leads to a statistical 
error larger by about $30\%$. The error on $\a_s(m_\t^2)$ has also
increased as a result of the larger errors on the updated HFLAV 2022 BFs, 
especially that of the $\pi^-3\pi^0$ mode. Clearly, a high-statistics 
Belle II determination of the two $4\p$ unit-normalized distributions 
would have the potential to significantly improve the error situation, as 
would improved determinations of the $4\pi$ BFs, especially that of the 
$\pi^-3\pi^0$ mode. In fact, already a high-statistics determination of the 
experimentally less challenging $2\p^-\p^+\p^0$ distribution would be of 
interest in this regard, as this mode constitutes about $80\%$ (by BF) of 
the sum of the two $4\p$-mode contributions. 

A further reason improved $\tau$-decay determinations of the 
$2\pi^-\pi^+\pi^0$- and $\pi^- 3\pi^0$-mode contributions to $\rho_{ud;V}(s)$ 
would be of interest is the connection to the $2\pi^+ 2\pi^-$ and 
$\pi^+\pi^- 2\pi^0$ contributions to the electromagnetic (EM) neutral vector, 
isovector-current spectral function, $\rho_{\rm EM}^{I=1}(s)$, which are 
determined by the corresponding electroproduction cross sections and play an 
important role in the dispersive determination of the hadronic vacuum 
polarization contribution to the anomalous magnetic moment of the muon. In 
the isospin limit, the Pais relations~\cite{pais4pi} provide expressions 
for the exclusive-mode $4\pi$ contributions to $\rho_{ud;V}(s)$ in terms 
of the exclusive-mode $4\pi$ contributions to $\rho_{\rm EM}^{I=1}(s)$, 
with the $\pi^-3\pi^0$ $\tau$ contribution determined by the $2\pi^+ 2\pi^-$ 
EM contribution, and the $2\pi^-\pi^+\pi^0$ $\tau$ contribution by a 
complementary linear combination of the EM $2\pi^+ 2\pi^-$ and 
$\pi^+\pi^- 2\pi^0$ contributions. The Pais-relation expectations obtained 
using publicly available exclusive-mode EM $4\pi$ results from the 
compilation of Ref.~\cite{KNT19}, have errors ranging from a factor of $\sim 2$ 
smaller than those of corresponding ALEPH and OPAL $\tau$ combination for
$s$ near $2$ GeV$^2$ to $\sim 5-6$ smaller near the $\tau$ kinematic endpoint. 
There are, however, signs of non-trivial tensions at a scale larger than 
those of the $\sim 1\%$ or so deviations expected due to isospin-breaking 
corrections to the Pais relations, between the central values and 
$s$-dependences of the $\tau$ data and the Pais-relation expectations in 
this same region, albeit with point-by-point significances limited by the 
significantly larger $\tau$ errors. Post-LEP-era improvements
to the $\tau$ $4\pi$ distributions would be of significant interest in 
this context.

Our new value for $\alpha_s$ represents the current best value 
obtained from the world's data for hadronic $\t$ decays, and supersedes 
the value of Ref.~\cite{alphas20}. The result is farther from the world 
average \cite{PDG} than that found in Ref.~\cite{alphas20}. The reason 
for the shift is the inclusion of new data, with the shift to a lower value 
being dominated by the change in the $4\p$ branching fractions.

The determination of $\a_s$ from $\t$ decays is kinematically limited by 
the $\t$ mass, and because of its relatively low value, non-perturbative 
effects in QCD have to be accounted for. Some, but not all, of the 
non-perturbative effects are captured by the OPE, as the OPE does not
account for the oscillatory resonance behavior clearly seen
in the spectral function at values of $s$ in the region between 2~GeV$^2$ 
and $m_\t^2$ ({\it cf.} Fig.~\ref{finalspec-fig}). Any determination of 
$\a_s$ from $\t$ decays thus needs to consider the possibility that
residual DV contributions may not be entirely numerically negligible in 
the FESRs used in that determination. We have investigated this issue using 
the {\it ansatz}, Eq.~(\ref{ansatz}), to describe violations of quark-hadron 
duality. The form of this {\it ansatz} is based on the presumed Regge-like 
behavior of the resonance spectrum at large $s$, combined with insights 
from large-$N_c$ QCD \cite{BCGMP}.\footnote{Further discussion of our 
assumptions can be found in Ref.~\cite{DVvstOPE}.} The multi-weight, 
multi-$s_0$ fits detailed in Sec.~\ref{results} provide non-trivial tests 
of this DV model since (i) the differently weighted theory integrals have 
different sensitivities to duality violations and (ii) integrated DV and 
non-perturbative OPE contributions have very different $s_0$ 
dependences.\footnote{For more tests also including OPAL or ALEPH
axial-vector data, we refer to Refs.~\cite{alphas2,alphas14}.} A new test is 
the sensitivity to a subleading correction represented by the $c/s$ term 
in Eq.~(\ref{ansatz}), shown in Fig.~\ref{ccorr}. Our systematic error estimate 
now takes into account the effect of such a possible subleading
correction, increasing the total error on our result for
$\alpha_s(m_\tau^2)$ by about $3\%$.   

The issue of duality violations, however, still needs to be studied 
in more detail. We first comment that, as more data for hadronic $\t$ 
decays become available in the future, our current analysis framework makes 
it straightforward to take these new data into account. Increased precision 
can help in testing the theoretical representations used on the theory side 
of the FESRs in the analysis more stringently. An alternate approach would 
be to employ recently developed techniques for obtaining weighted integrals 
of the spectral function from lattice QCD, using, for example, the method 
of Ref.~\cite{taulatQCD}. A major advantage of the alternate approach is the 
absence of the kinematic restriction to $s_0\leq m_\t^2$ which is 
unavoidable for analyses based on experimental data of the type employed 
in this paper. We plan to investigate this approach in the near future.

\vspace{3ex}
\noindent {\bf Acknowledgments}
\vspace{3ex}

We thank Hisaki Hayashii for discussion of the Belle data, 
Sven Menke for discussion of the OPAL data, and Jon Urheim for
discussion of the CLEO data.  We also thank Mattia Bruno for
discussion of Ref.~\cite{BS}. 
LMM would like to thank the Department of Physics and Astronomy at
San Francisco State University for hospitality. DB's work was supported 
by the S\~ao Paulo Research Foundation (FAPESP) grant No. 2021/06756-6 and 
by CNPq grant No. 308979/2021-4. AE and MG are supported by the 
U.S.\ Department of Energy, Office of Science, Office of High Energy 
Physics, under Award No. DE-SC0013682. KM is supported by a grant from the 
Natural Sciences and Engineering Research Council of Canada.
LMM is supported by FAPESP grants No.~2023/08482-6 and No.~2022/16553-8.
SP is supported by the Spanish Ministerio de Ciencia e Innovacion,
grants PID2020-112965GB-I00 and PID2023-146142NB-I00,
 and by the Departament de Recerca i Universities from Generalitat de 
Catalunya to the Grup de Recerca 00649 (Codi: 2021 SGR 00649).
IFAE is partially funded by the CERCA program of the Generalitat de Catalunya.

\appendix
\section{\label{Abiascorr} Approximate d'Agostini bias correction}
In this Appendix, we review briefly how global normalization constants,
with their associated uncertainties, can lead to a d'Agostini bias
in the fitted results \cite{Abias} for the $4\pi$-mode data
combination. In addition, we present the adaptation of the NNPDF
procedure of Ref.~\cite{NNPDF} employed in our $4\pi$-mode combination
to approximately avoid this bias.

As explained in Sec.~\ref{4picomb}, it turns out to be possible to
combine only the separate, already summed, ALEPH and OPAL
$2\p^-\p^+\p^0$ plus $\p^-3\p^0$ mode contribution totals. This
implies that we need to multiply the kinematic-weight-rescaled
exclusive-mode unit-normalized distributions from ALEPH and OPAL
by the corresponding BFs (divided by $\cf$) before carrying out
the fit combining all $4\p$ data. As these BFs have non-zero
errors, the combination suffers from a potential d'Agostini bias.

If we wish to combine the spectral data for a certain exclusive
mode for two experiments $k$ and $\ell$, we start by
constructing the exclusive-mode spectral functions $\r^{k,\ell}_i$
from the kinematic-weight-rescaled unit-normalized distributions
$u^{k,\ell}_i$. This involves multiplying by the BF $B$ and the
factor $1/\cf$ with $\cf$ given in Eq.~(\ref{C}) (the index $i$ labels the
bins for each experiment):
\begin{equation}
\label{rho}
\r^k_i=\frac{B}{\cf}\,u^k_i\ .
\end{equation}
With $c^k_{ij}$ the covariance matrix for the distributions $u^k_i$,
the covariance matrix for $\r^k$ is given by
\begin{equation}
\label{covrho}
\mbox{cov}^{kk}_{ij}=\frac{B^2}{\cf^2}\,c^k_{ij}
+\frac{1}{\cf^2}\,u^k_i u^k_j\,\s_B^2+\frac{B^2}{\cf^4}
\,u^k_i u^k_j\,\s_{\cf}^2\ ,
\end{equation}
where $\s_B$ is the error on $B$ and $\s_{\cf}$ the error on $\cf$.\footnote{We
neglect correlations between $u^k_i$, $B$ and $\cf$.}
The errors on $B$ and $\cf$ also induce covariances between
data for the same mode from two different experiments $k$ and $\ell$
of the form
\begin{equation}
\label{mncov}
\mbox{cov}^{k\ell}_{ij}=\frac{1}{\cf^2}\,u^k_i u^\ell_j\,\s_B^2
+\frac{B^2}{\cf^4}\,u^k_i u^\ell_j\,\s_{\cf}^2\ .
\end{equation}
The presence of terms in Eq.~(\ref{covrho}) and Eq.~(\ref{mncov}) that depend on
$u^k_i u^k_j$ and $u^k_i u^\ell_j$ leads to a potential d'Agostini bias 
when we combine the spectral functions $\r^k_i$ for the two $4\p$ modes:
data points with higher values of the distribution $u^k_i$ have
less impact on the determination of the $\rho^{(m)}(s)$ cluster results
in the fitted combination. In other words, this bias leads to our combined
spectral function, $\rho^{(m)}(s)$ having lower central values
than one would have obtained if $\s_B$ and $\s_{\cf}$ were to be
negligibly small \cite{NNPDF,Abias}.

\begin{figure}[t]
\vspace*{4ex}
\begin{center}
\includegraphics*[width=15.25cm]{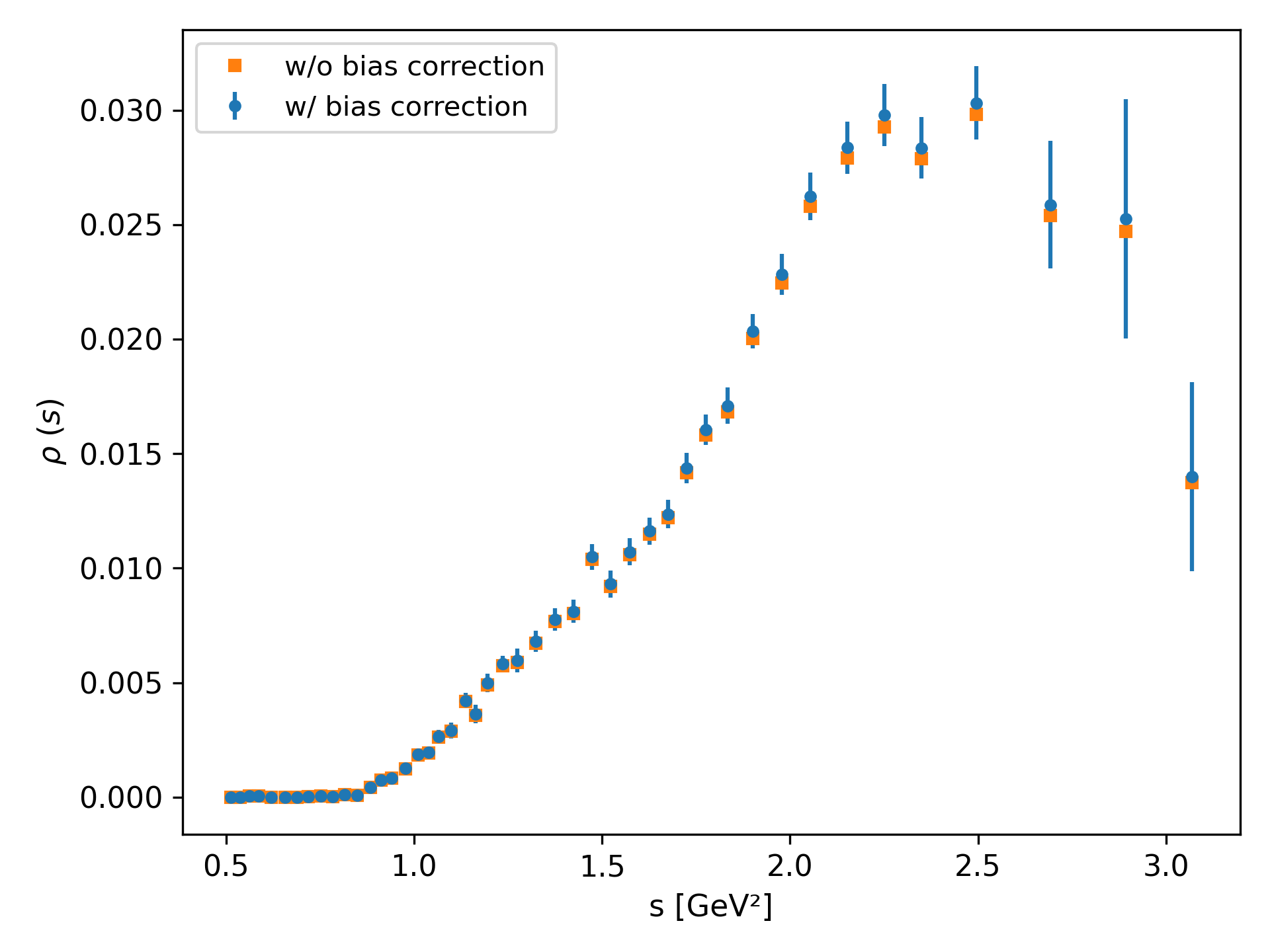}
\end{center}
\begin{quotation}
\floatcaption{bias-fig}%
{{\it The $4\p$ combined spectral function, with (blue points) and without 
(orange points) bias correction.}}
\end{quotation}
\vspace*{-4ex}
\end{figure}

In Ref.~\cite{NNPDF}, an iterative procedure was proposed to minimize the
effect of the d'Agostini bias. The idea is to iterate the fit by, at
each step of the iteration, replacing the original data $u^m_i$ in
Eqs.~(\ref{covrho}) and~(\ref{mncov}) by the corresponding result from the
previous iteration of the fit, \ie, by $\r^{(m)}(s)/(B/\cf)$, interpolated
to the bins of the experiments $k$ and $\ell$, redoing the fit with
this as updated input. As explained in detail in Ref.~\cite{NNPDF}, this
iterative procedure is expected to converge rapidly.

For the $2\p$ mode, the bias problem can be avoided by directly combining 
unit-normalized spectral data from different experiments (in our case, 
ALEPH, OPAL and Belle). However, the ill-behaved covariance
matrices for the $2\p^-\p^+\p_0$ and $\p^-3\p^0$ modes force us
to first form the full two-mode $4\pi$ sums,
$\rho_{4\pi}(s)=\rho_{2\pi^{-}\pi^{+}\pi^{0}}(s)
+\rho_{\p^-3\pi^{0}}(s)$, for each of ALEPH and OPAL, and only then
combine the data from the two experiments. It follows that after the 
first step in the iteration we have only the combined ``clustered'' 
spectral function $\left[\r_{4\pi}\right]^{(m)}(s)$, and no longer 
have information about the precise contribution from the individual 
$2\pi^{-}\pi^{+}\pi^{0}$ and $\pi^{-}3\pi^{0}$ modes. In order to 
iterate the fit we use the following approximate procedure. We first 
interpolate $\left[\r_{4\pi}\right]^{(m)}(s)$ to the ALEPH and OPAL bins, 
thus constructing  ``improved'' ALEPH and OPAL $4\p$ spectral distributions.
Then, in order to turn this into ``improved'' ALEPH and OPAL $2\p^-\p^+\p^0$ 
and $\p^-3\p^0$ distributions, we split these $4\p$ distributions according 
to the bin-by-bin ratios for these two modes for each of the two experiments. 
This then allows us to carry out the iterative procedure proposed by
Ref.~\cite{NNPDF}. We find convergence after two steps. This procedure is
approximate, because it rests on the assumption that keeping the ratios
of the original ALEPH and OPAL experimental data between the two modes
the same during the iteration procedure is a reasonable approximation. 
We note that these ratios are not exactly the same for ALEPH and OPAL. 

\begin{figure}[t]
\vspace*{4ex}
\begin{center}
\includegraphics*[width=15.25cm]{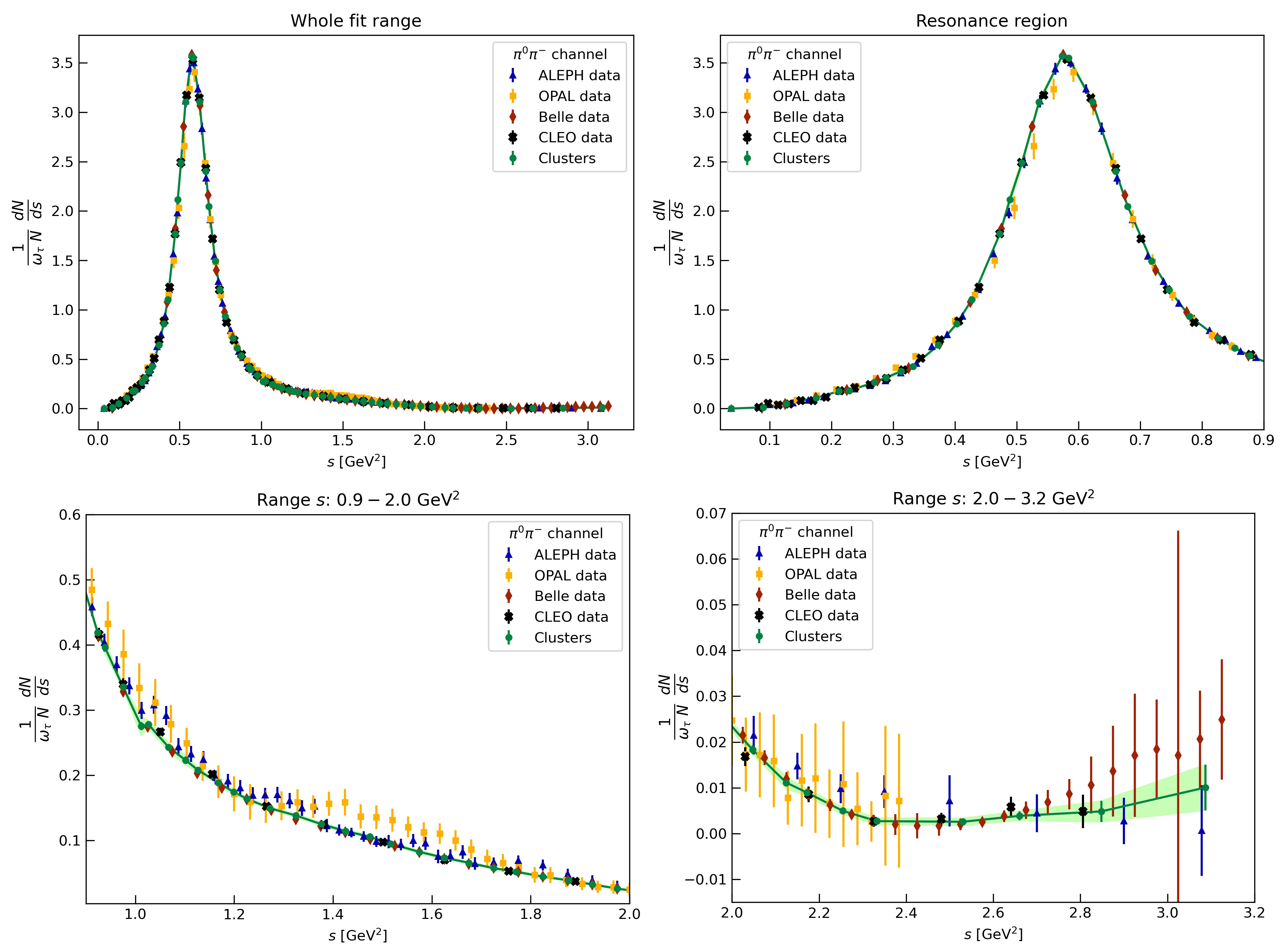}
\end{center}
\begin{quotation}
\floatcaption{AOBC-fig}%
{{\it The $2\p$ unit-normalized number distribution combination, divided by 
$w_T(s)$, together with the data from the individual ALEPH, OPAL,
Belle and CLEO experiments. The error bars represent the 
uninflated errors, while inflated errors are represented by the green band.}}
\end{quotation}
\vspace*{-4ex}
\end{figure}

\begin{figure}[t!]
\vspace*{4ex}
\begin{center}
\includegraphics*[width=15.25cm]{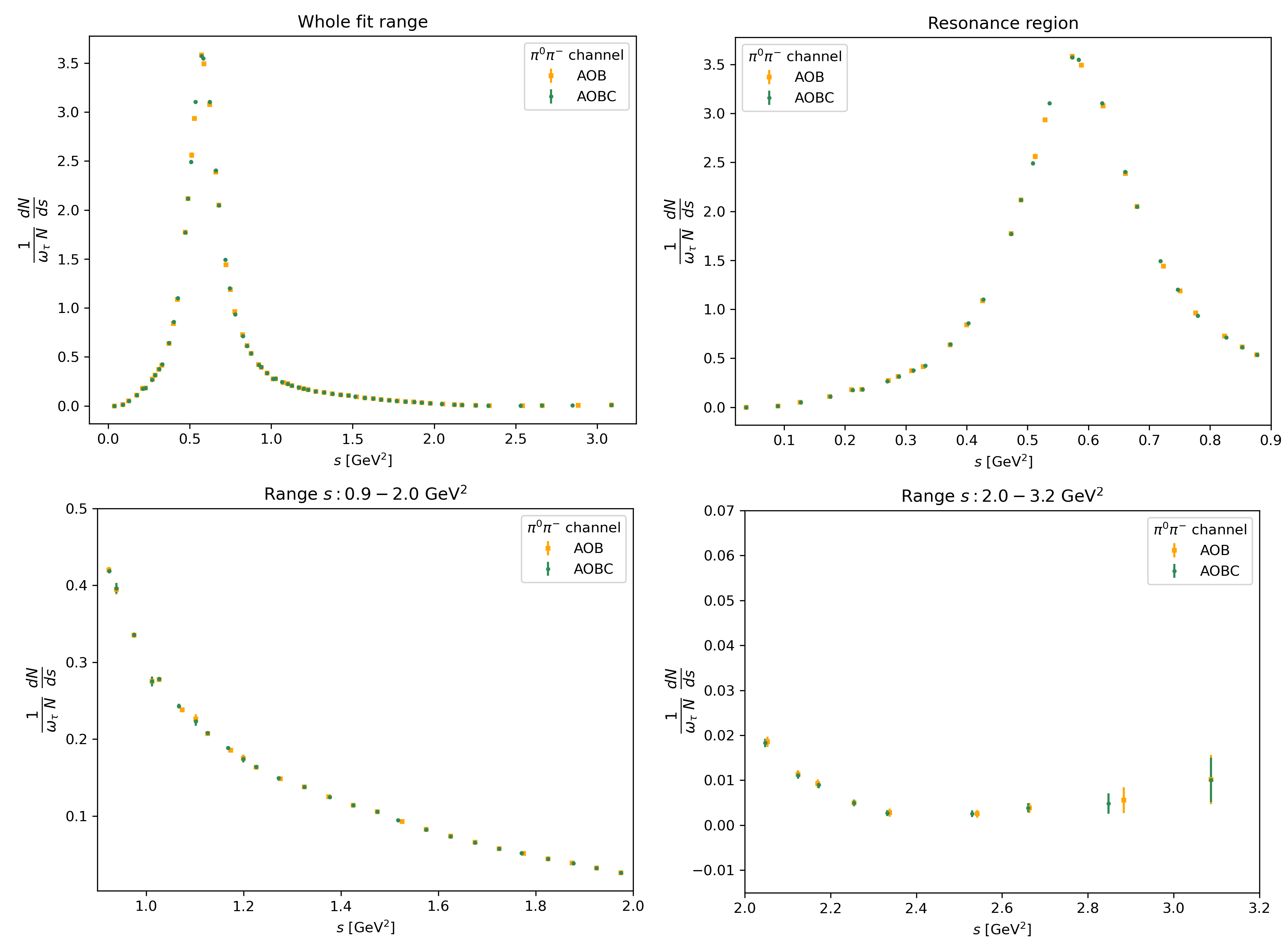}
\end{center}
\begin{quotation}
\floatcaption{2picombwithCLEO-fig}%
{{\it Comparison of the $2\p$ unit-normalized number distribution 
combination(divided by $w_T(s)$) with orange points indicating the 
ALEPH, OPAL and Belle combination (AOB) of Sec.~\ref{datasec}, 
and green points indicating the ALEPH, OPAL, Belle and CLEO 
combination (AOBC).}}
\end{quotation}
\vspace*{-4ex}
\end{figure}

\section{\label{CLEOsec} \begin{boldmath}Inclusion of the CLEO $\p^-\p^0$-mode
data\end{boldmath}}
In this appendix, we investigate the effect of including the CLEO data for the 
$\t\to\p^-\p^0\n_\t$ decay mode \cite{CLEO,CLEO2pi} in the construction of the 
combined $2\p$ spectral function.

For this data set, only statistical errors, but not systematic errors, 
are available on a bin-by-bin basis. This raises the question as to 
whether the CLEO data can be reliably combined with the other three data 
sets, for which both statistical and systematic covariance information is 
available. In this appendix, we nevertheless include the CLEO data set in 
a four-experiment combination of the $\p^-\p^0$ mode, using only the 
statistical error information available for CLEO, to see to what extent 
the CLEO data modify the combination. Starting from the unit-normalized 
number distributions from all four experiments, we follow the same strategy as
used in Sec.~\ref{2picomb}. The resulting unit-normalized number distribution, 
divided by the kinematic weight $w_T(s)$, is shown in Fig.~\ref{AOBC-fig}.
We use the same number of clusters, 63, and assign the CLEO data points
to the clusters with the nearest invariant-mass-squared value. Since they are
determined from Eq.~(\ref{clusters}), the cluster $s$ values for the new 
combination are not identical to those for the combination of 
Sec.~\ref{2picomb}, but they are close, as can be seen in 
Fig.~\ref{2picombwithCLEO-fig}. This figure shows the $\p^-\p^0$ 
unit-normalized number distribution combination, divided by $w_T(s)$, 
with (AOBC, green points) and without (AOB, orange points) the CLEO data.
From the figure, it can be seen that the CLEO data have a rather limited 
effect on the combined $2\p$ spectrum. The global fit is of good quality, 
with $\c^2=201$ for 193 degrees of freedom, and a $p$-value of 
0.34. The local $p$-values per cluster are very similar to those 
shown in Fig.~\ref{2pipvalue-fig}.

We have also investigated the effect of including the CLEO data in our 
inclusive combined non-strange $V$ spectral function on the values for 
$\a_s(m_\t^2)$ and the DV parameters. It turns out that the effect is very 
small. As an example, consider the fit to $I_{\rm ex}^{(w_0)}(s_0)$
with $s_0^{\rm min}=1.5747$~GeV$^2$, which has a $p$-value of 0.84
and leads to the parameter values
\begin{eqnarray}
\label{w0parsC}
\a_s(m_\t^2)&=&0.2997(96)\ ,\\
\d&=&3.04(39)\ ,\nonumber\\
\g&=&0.85(24)\ \mbox{GeV}^{-2}\ ,\nonumber\\
\a&=&-1.30(73)\ ,\nonumber\\
\b&=&3.76(38)\ \mbox{GeV}^{-2}\ .\nonumber
\end{eqnarray}
Both central values and errors are very close to the corresponding entries
in Table~\ref{tabw0}. In particular, we see that inclusion of the CLEO data 
does not increase the precision that can be obtained from the fit to the 
combined spectral function. This remains true for the parameter values 
obtained with the other $s_0^{\rm min}$ values in the range used to produce 
our result~(\ref{w0pars}).   

Given these results, and given the fact that no complete covariance 
information is available for the CLEO $\p^-\p^0$ data set, we do not 
include the CLEO data in producing our central results, Table~\ref{tabspec} 
for the inclusive vector isovector spectral function, and Eq.~(\ref{final})
for the value of $\a_s(m_\t^2)$.

\begin{boldmath}
\section{\label{2pi4pi} Combination of $2\p$ and $4\p$ spectral distributions}
\end{boldmath}
We begin by grouping the data together into a single vector \cite{Helene2006},
\begin{equation}
\label{data}
d_0=(u_{2\p,i},u_{1\p^0,i},u_{3\p^0,i},B_{2\p},B_{1\p^0},B_{3\p^0},\cf)\ ,
\end{equation}
where $u_X$ indicates a
kinematic-weight-rescaled unit-normalized spectral
distribution, and $X=2\p$, $1\p^0$ and $3\p^0$ are short forms for
the $\p^-\p^0$, $2\p^-\p^+\p^0$ and $\p^-3\p^0$ exclusive modes,
respectively. The $B_X$ are the corresponding BFs given in Eq.~(\ref{BX}),
$\cf$ is the combination of external constants given in Eq.~(\ref{C}), and
$u_{2\p}$ is the full combined $2\p$ kinematic-weight-rescaled version of
the unit-normalized $2\pi$ distribution obtained in Sec.~\ref{2picomb}.
The quantities $u_{1\p^0,i}$ and $u_{3\p^0,i}$, in contrast,
are the unions, for each of the two $4\p$ exclusive modes, of the ALEPH
and OPAL kinematic-weight-rescaled unit-normalized spectral distribution
data, {\it before} performing the two-mode combination described
in Sec.~\ref{4picomb}. If the total numbers of ALEPH and OPAL $1\p^0$
and $3\p^0$ data points are $N_1$ and $N_3$, respectively, the length
of $d_0$ is $63+N_1+N_3+3+1$. The covariance matrix $C_0$ for Eq.~(\ref{data})
is block diagonal, containing the blocks for the $2\p$ combined
$63\times 63$ covariance matrix $\cc_{2\p}$, the merged ALEPH and
OPAL $N_1\times N_1$ $1\p^0$ and $N_3\times N_3$ $3\p^0$ covariance
matrices as in Sec.~\ref{4picomb}, the $3\times 3$ covariance matrix for
the BFs constructed from Eq.~(\ref{BX}), and the squared-error for $\cf$ in
the last diagonal entry.\footnote{Because the relative error on $B_e$
is very small ($O(0.1\%)$) we can neglect the correlations between $B_e$
and the hadronic BFs.}

These data can be combined into the spectral data vector
\begin{equation}
\label{spectralvec}
d_1=\frac{1}{\cf}\left(B_{2\p}u_{2\p,i},B_{1\p^0}u_{1\p^0,i},
B_{3\p^0}u_{3\p^0,i}\right)\ .
\end{equation}
With $D_1$ the rectangular matrix of derivatives of the entries of
Eq.~(\ref{spectralvec}) with respect to those of Eq.~(\ref{data}),
\begin{equation}
\label{deriv}
D_1=\frac{1}{\cf}\left(\begin{array}{ccccccc}
B_{2\p}\d_{ij} & 0 & 0 & u_{2\p,i} & 0 & 0 & -B_{2\p}u_{2\p,i}/\cf \\
0 & B_{1\p^0}\d_{ij} & 0 & 0 & u_{1\p^0,i}  & 0 & -B_{1\p^0}u_{1\p^0,i}/\cf\\
0 & 0 & B_{3\p^0}\d_{ij} & 0 & 0 & u_{3\p^0,i} & -B_{3\p^0}u_{3\p^0,i}/\cf
\end{array}\right)\ ,
\end{equation}
the $(63+N_1+N_3)\times (63+N_1+N_3)$ covariance matrix $C_1$ for
$d_1$ is given by
\begin{equation}
\label{C1}
C_1=D_1C_0D_1^T\ .
\end{equation}
We note that $C_1$ is no longer block-diagonal, because of the form of $D_1$.

In the next step, we apply a transformation to $d_1$ to arrive at
\begin{equation}
\label{d2}
d_2=\left(\begin{array}{c}\r_{2\p}\\ \r_{4\p}\end{array}\right)\ ,
\end{equation}
where $\r_{2\p}=B_{2\p}u_{2\p}/\cf$ is the $2\p$ spectral function
obtained in Sec.~\ref{2picomb}, \ie, the first entry of $d_1$, and
\begin{equation}
\label{rho4pi}
\r_{4\p}=M_{4\p}\,\frac{1}{\cf}\left(\begin{array}{c}
B_{1\p^0}u_{1\p^0,i}\\B_{3\p^0}u_{3\p^0,i}
\end{array}\right)\ ,
\end{equation}
with $M_{4\p}$ the $46\times (N_1+N_3)$ rectangular matrix,
defined in Eq.~(\ref{W2fit}), which combines the ALEPH and OPAL $4\p$
data into the 46-cluster combined two-mode $4\p$ spectral
function sum. We thus have that
\begin{equation}
\label{d1tod2}
d_2=D_2 d_1^T
\end{equation}
(the transpose appears because we gave $d_1$ as a row vector),
with
\begin{equation}
\label{D2}
D_2=\left(\begin{array}{cc}{\bf 1} & 0\\ 0 & M_{4\p}\end{array}\right)\ ,
\end{equation}
in which ${\bf 1}$ is the $63\times 63$ unit matrix. The covariance matrix
for $d_2$ is
\begin{equation}
\label{C2}
C_2=D_2C_1D_2^T\ .
\end{equation}
Finally, we linearly interpolate the $4\p$ spectral function $\r_{4\p}$
to the $s$ values of the 63 $2\p$ clusters, and add the interpolated
$\r_{4\p}$ to $\r_{2\p}$. The covariances of the resulting
$2\p+4\p$ spectral function, $\r_{2\pi +4\pi}$, are straightforwardly
obtainable from $C_2$ and the coefficients used to interpolate the
$4\pi$ results from the $s$ values of the 46-cluster $4\pi$ $s$ set
to those of the 63-cluster $2\pi$ set. We emphasize that the matrix
$C_0$ contains all data covariances, and that, when parts of it
are modified in the fit of Sec.~\ref{4picomb}; all correlations
are retained in the error propagation.


\end{document}